\documentclass[letter,11pt]{article}
\pdfoutput=1 % if your are submitting a pdflatex (i.e. if you have
\usepackage{jcappub} % for details on the use of the package, please
\usepackage[T1]{fontenc} % if needed
\usepackage{ dsfont }
\usepackage{amsmath,amssymb,calc}
\usepackage{color}
\usepackage{amssymb}
\usepackage{graphicx, epsfig, bm}
\usepackage{hyperref}
\usepackage{soul}
 \usepackage{multicol}
\usepackage{changepage}
\usepackage{subfig}

\usepackage{color}
\usepackage{ifthen}

\def\nn{\nonumber}
\def\({\left(}
\def\){\right)}
\def\[{\left[}
\def\]{\right]}

\def\exd{{\hbox{d}}}
\def\d{\exd}

\def\nn{\nonumber}
\def\({\left(}
\def\){\right)}
\def\[{\left[}
\def\]{\right]}

 \newcommand{\n}{\nonumber\\}
  \newcommand\ddfrac[2]{\frac{\displaystyle #1}{\displaystyle #2}}
\newcommand{\beq}{\begin{equation}}
\newcommand{\beqn}{\begin{eqnarray}}
\newcommand{\eeq}{\end{equation}}
\newcommand{\eeqn}{\end{eqnarray}}
  \newcommand{\be}{\begin{equation}}
  \newcommand{\ee}{\end{equation}}
  \newcommand{\bea}{\begin{eqnarray}}
\newcommand{\eea}{\end{eqnarray}}

\usepackage{color}

\bibstyle{JHEP}

\title{Anisotropic Massive Gauge-flation}

\author{Peter Adshead and}
\author{Aike Liu}

\affiliation{Department of Physics, University of Illinois at Urbana-Champaign, Urbana, Illinois 61801, U.S.A.}

\emailAdd{adshead@illinois.edu}
\emailAdd{aikeliu2@illinois.edu}

\abstract{We study  anisotropic inflationary solutions in massive Gauge-flation. We work with the theory in both the Stueckelberg and dynamical symmetry-breaking limits and demonstrate that extended periods of accelerated anisotropic expansion are possible. In the case of dynamical symmetry breaking, we show that spacetime can transition from isotropic quasi-de Sitter space  to an accelerating Bianchi spacetime due to a rolling Higgs field --- the spacetime can develop hair. Similarly, symmetry restoring transitions are possible from accelerating Bianchi spacetime to quasi-de Sitter space   --- the spacetime can lose its hair. These transitions can be arranged to occur quickly, within an $e$-folding or so, or over tens of $e$-folds.}

\begin{document}
\maketitle
\flushbottom

\section{Introduction}

The inflationary paradigm --- a period of early accelerated expansion \cite{Guth:1980zm,Starobinsky:1980te,Sato:1980yn, Linde:1981mu,Albrecht:1982wi,Linde:1983gd} --- is in good shape. Inflation solves classic problems associated with the initial conditions for the  hot Big Bang, while simultaneously generating the seeds for the subsequent gravitational growth of structure  via quantum mechanical fluctuations of the fields and metric  \cite{Lukash:1980iv,Press:1980zz,Mukhanov:1981xt,Guth:1982ec,Hawking:1982cz,Bardeen:1983qw}. Increasingly accurate measurements of the cosmic microwave background (CMB) \cite{Komatsu:2008hk,Ade:2015xua} have revealed the initial density fluctuations to be red-tilted and gaussian \cite{Ade:2013ydc, Ade:2015ava}, in accord with generic inflationary predictions. However, the simplest large-field single-field slow-roll models (e.g.\ \cite{Linde:1983gd, Freese:1990rb})  are now disfavored   by data due to constraints on the tensor-to-scalar ratio, $r < 0.07$ at 95\% \cite{Ade:2015tva, Ade:2015lrj, Array:2015xqh}. Future CMB measurements will potentially constrain (or measure) the tensor-to-scalar ratio to $\sigma_r \sim 10^{-3}$  \cite{Abazajian:2016yjj} and motivate the study models of inflation that can populate this region of parameter space.

The large-scale isotropy of our Universe is generally thought to be a generic prediction of inflation due to the cosmic no-hair conjecture  \cite{Wald:1983ky}. However, there are hints  of statistical anisotropy on very large scales in the CMB, from the apparent alignment of the low-$\ell$ multipoles  \cite{Tegmark:2003ve,deOliveira-Costa:2003utu, Copi:2005ff} (dubbed the `axis of evil' \cite{Land:2005ad}) to evidence of a hemispherical  power asymmetry and dipolar anomaly \cite{Eriksen:2003db, Hansen:2004vq,  Eriksen:2007pc, Hansen:2008ym, Groeneboom:2008fz, 2009ApJ...699..985H, Ade:2013nlj, Ade:2015hxq}. The significance of these anomalies is disputed \cite{Rassat:2014yna} and several studies have placed very tight limits on the allowed statistical anisotropy  \cite{Kim:2013gka, Ade:2015lrj, Ramazanov:2013wea, Ramazanov:2016gjl}.  Theorists have attempted to explain these anomalies in a variety of ways using inflationary dynamics. In particular, the gradient effects of large, superhorizon fluctuations produced during inflation \cite{Gordon:2005ai, Gordon:2006ag,  Donoghue:2007ze,Erickcek:2008sm, Erickcek:2008jp} as well as imprints of an earlier anisotropic phase \cite{Gumrukcuoglu:2007bx} have been postulated to be responsible. The dynamics of fixed-norm vector fields  \cite{Ackerman:2007nb} and non-minimally coupled massive vectors \cite{Kanno:2008gn} were demonstrated to generate periods of extended anisotropic inflation and associated anisotropic fluctuations. However, these models were later shown to have pathological instabilities \cite{Bluhm:2008yt, Himmetoglu:2008zp, Himmetoglu:2008hx}. After the construction of stable inflationary models that use vector fields to generate anisotropic inflation \cite{Watanabe:2009ct}, there has been a surge of interest in anisotropic inflationary scenarios \cite{Watanabe:2010bu,Murata:2011wv,Soda:2012zm,Ohashi:2013mka,Ohashi:2013qba,Ito:2015sxj,Ito:2016aai,Ito:2017bnn, Muller:2017nxg,Fujita:2017lfu,Do:2017onf}. On the other hand, it is shown recently that anisotropic inflation can also be realized by modified gravity theories without introducing vector fields \cite{Tahara:2018orv}. These models  provide explicit counter-examples to the cosmic no-hair conjecture.

In this work, we investigate anisotropic inflation in models of inflation built from non-Abelian gauge fields --- Gauge-flation \cite{Maleknejad:2011sq,Maleknejad:2011jw}, and Chromo-Natural inflation \cite{Adshead:2012kp}. Gauge-flation is a model of inflation that does not contain any scalar fields. Instead, non-Abelian gauge fields in an isotropic, flavor-space locked configuration generate an inflationary phase via an $(F\tilde{F})^2$ term. In Chromo-Natural inflation, axion inflation on a steep potential is facilitated by a Chern-Simons interaction with a flavor-space locked non-Abelian gauge field. The models are related by integrating out the axion about the minima of a quadratic potential \cite{Adshead:2012qe, SheikhJabbari:2012qf}. The rotation symmetry, which is usually broken by the vector nature of the gauge fields, is restored in these models by identifying the SU(2) group with the spatial symmetry group SO(3). While the original models of Gauge-flation and Chromo-Natural models are ruled out \cite{Adshead:2013qp, Adshead:2013nka, Namba:2013kia}, they can be brought into agreement with CMB data by introducing masses for the gauge fields \cite{Nieto:2016gnp, Adshead:2016omu,Adshead:2017hnc}, or by flattening the Chromo-Natural inflation potential  \cite{Caldwell:2017chz}.

Maleknejad et.\ al.\ have demonstrated that an axially symmetric configuration in a Bianchi spacetime  evolves to an isotropic universe within a few $e$-folds, regardless of initial conditions for both Gauge-flation \cite{Maleknejad:2011jr} and Chromo-Natural inflation \cite{Maleknejad:2013npa}. This isotropizing behavior is consistent with the no-hair conjecture.  In this work, we point out that the massive versions of both Gauge-flation and Chromo-Natural inflation allow for other possibilities due to the symmetry breaking. Because the gauge symmetries of the vector fields are identified with spatial rotations, different symmetry breaking patterns can give rise to different spacetime symmetry patterns. In this paper, we explore the breaking of isotropy by the gauge-field mass terms. We begin by examining the theory in the broken phase in which the gauge field masses are non-dynamical and are introduced by the Stueckelberg mechanism. We then study dynamical symmetry breaking by a Higgs sector where an initial isotropic, massless scenario evolves dynamically to a broken phase \emph{during} inflation. We also consider the case where an initially massive, anisotropic configuration evolves to a  massless, symmetry restoring state. We find that massive Gauge-flation can generically generate extended periods of anisotropic inflation. We also find that an initially spherically symmetric quasi-de Sitter spacetime can develop hair  by dynamically evolving to  an accelerating Bianchi spacetime from  a quasi-de Sitter spacetime, and vice-versa. The magnitude of the resulting anisotropy is similar to that found in anisotropic power-law inflation presented in \cite{Do:2017onf, Ohashi:2013pca}.

This paper is organized as follows. In section \ref{sec-review} we review the original massless \cite{Maleknejad:2011sq} and massive \cite{Nieto:2016gnp, Adshead:2016omu} Gauge-flation models. In section \ref{sec-static} we study massive Gauge-flation in an axially symmetric spacetime and explore the effects of varying the gauge field masses. In section \ref{sec:dynamical}, we allow the vacuum expectation value of the Higgs field (which gives masses to the gauge field) to dynamically evolve on a potential and study the dynamical breaking of the gauge symmetry during inflation. We conclude in section \ref{sec:conclusions}.  Throughout, we work in natural units where $\hbar = c = M_{\rm Pl} = 1$.

\section{Massive and massless Gauge-flation}\label{sec-review}

We begin by briefly reviewing the theory of Gauge-flation \cite{Maleknejad:2011jw, Maleknejad:2011sq} and massive Gauge-flation  \cite{Nieto:2016gnp,Adshead:2017hnc} in a Friedmann-Robertson-Walker (FRW) spacetime. Massive \cite{Nieto:2016gnp}, or Higgsed Gauge-flation  \cite{Adshead:2017hnc},  is described by the action 
\begin{align}\label{eqn:HiggsGF}
\mathcal{S} =  \int \exd^4x\sqrt{-{g}}\left[\frac{R}{2} -\frac{1}{4}F^a_{~\mu\nu}F^{a\, \mu\nu}+\frac{\kappa}{384
}(\epsilon^{\mu\nu\lambda\sigma}F^a_{~\mu\nu}F^a_{~\lambda\sigma})^2 - \frac{1}{2}(D\Psi)^2- V(\Psi) \right],
\end{align}
where $\Psi$ is a scalar field (Higgs multiplet), $A^a_{~\mu}$ are SU(2) gauge fields, with the field strength tensor defined\footnote{Our convention for the antisymmetric tensor is $\epsilon^{0123} = 1/\sqrt{-g}$,  while our spacetime metric signature is $(-,+,+,+)$.  Here and throughout, Greek letters  denote spacetime indices, Roman letters from the start of the alphabet denote gauge indices and Roman letters from the middle of the alphabet denote spatial indices. }
\be
F^a_{~\mu\nu} = \partial_\mu A^a_{~\nu}-\partial_\nu A^a_{~\mu}+g\epsilon^a_{~bc}A^b_{~\mu}A^c_{~\nu}\ .
\ee

We initially assume that the theory is in the broken phase, where the Higgs has a static vev: $\langle \Psi \rangle \neq 0$ with $V(\langle \Psi \rangle ) = 0$ and $d{\langle \Psi \rangle}/dt = 0$. Without loss of generality we take the action to be in the form\footnote{In the notation of \cite{Adshead:2017hnc}, $m_a = g Z_0$.}  
\begin{align}\label{eqn:GFaction22} 
\mathcal{S} =  \int \exd^4x\sqrt{-{g}}\left[\frac{ R}{2}-\frac{1}{4}F^a_{~\mu\nu}F^{a\, \mu\nu}+\frac{\kappa}{384
}(\epsilon^{\mu\nu\lambda\sigma}F^a_{~\mu\nu}F^a_{~\lambda\sigma})^2- \frac{1}{2}\sum_{a = 1}^{3} m^{~2}_aA^{a}_{~\mu} A^{a\,\mu} \right].
\end{align}
In writing the action in the form of eq.\ \eqref{eqn:GFaction22}, we have made an SU(2) rotation to diagonalize the mass matrix.

%%%%%%%%%%%%%%%%%%%%%%%%%%%%%%%%%%
%
\subsection{Isotropic solutions}\label{sec:backgroundeqns}
%
%%%%%%%%%%%%%%%%%%%%%%%%%%%%%%%%%%

Gauge-flation is obtained by putting the gauge fields in the classical, flavor-space locked configuration 
\begin{align}\label{eqn:gaugevev}
A^a_{0} = & 0, \quad A^a_{i} =  \psi\, e^a_{~i} = a\psi \, \delta^{a}_{i}, \quad a= e^\alpha, 
\end{align}
where $e^a_{~i}$ is the spatial tetrad that identifies the gauge index $a$ with spatial index $i$. We denote $\phi = a\psi$ in the following discussion, where $a$ is the isotropic scale factor.\footnote{We use $a$ to denote both the gauge index as well as the scale factor, and distinguishing between them should be fairly obvious from the context.} This configuration ensures that the resulting stress tensor is consistent with the symmetries of FRW spacetime. We write the metric in FRW form
\begin{align}
\d s^2 = - N^2 \d t^2 + a^2 \delta_{ij}\d x^i \d x^j,
\end{align}
where $N$ is the lapse function, and $N = 1$ on the background.

For the gauge field configuration in eq.\ \eqref{eqn:gaugevev}, the components of field-strength tensor are
\begin{align}
F^a_{0i} = & { \dot{\phi}} \, \delta^{a}_{i},\quad
F^a_{ij}  =  g \phi^2 f^{a}_{ij}, 
\end{align}
where $f^a_{ij}$ the structure constants of SU(2). Here and throughout,  an overdot  represents a derivative with respect to cosmic time, while a prime represents a derivative with respect to conformal time.  For these degrees of freedom, the reduced action takes the form\footnote{Note that the self consistency of the isotropic solutions requires that $m_1 = m_2 = m_3 \equiv m$.}
\begin{align}\label{eqn:minisupact}
\mathcal{L} =  a^{3}N \Bigg[-3 \frac{\dot a^2}{a^2 N^2} + \frac{3}{2}\frac{\dot \phi^2}{a^2 N^2}  - \frac{3}{2}g^{2}\frac{\phi^4}{a^4} + \frac{3}{2 N^2}\kappa \frac{ g^2\phi^4\dot{\phi}^2}{a^6}-\frac{3}{2}m^2 \frac{\phi^2}{a^2} \Bigg]\, .
\end{align}
It is convenient to separate the contributions from Yang-Mills action, the $\kappa (F\tilde{F})^2$ term, and the Higgs sector,
\begin{align}
\mathcal{L} _{\rm red} = \mathcal{L} ^{\rm YM}+ \mathcal{L} ^{\kappa}+\mathcal{L} ^{m} ,\quad
T_{\mu\nu} = T_{\mu\nu}^{\rm YM} + T_{\mu\nu}^{\kappa}+T_{\mu\nu}^m.
\end{align}

The Yang-Mills action contributes terms to the stress tensor that have the same equation of state as radiation, i.e. $P_{\rm YM} = \rho_{\rm YM}/3$, where
\begin{align}
\rho_{\rm YM} =\frac{3}{2}\frac{\dot \phi^2}{a^2}  + \frac{3}{2}g^{2}\frac{\phi^4}{a^4},
\end{align}
while the terms result from the $\kappa$ term give the equation of state of a cosmological constant. That is $P_{\kappa} = -\rho_{\kappa}$, where
\begin{align}
\rho_{\kappa} = & \frac{3\kappa}{2}\frac{g^2 \dot\phi^2\phi^4}{a^6}.
\end{align}
As long as $\rho_{\kappa} \gg \rho_{\rm YM}$, the background spacetime undergoes a phase of accelerated expansion.

The Higgs sector yields additional contributions to both the energy density and the pressure,
\begin{align}
\rho_{m} = \frac{3}{2}m^2 \frac{\phi^2}{a^2} , \quad P_{m} =- \frac{1}{2}m^2\frac{\phi^2}{a^2}.
\end{align}
Note the equation of state for these terms is $w = -1/3$, and thus the presence of the symmetry-breaking sector does not affect the condition for accelerated expansion, which remains  $\rho_{\rm YM} \ll \rho_{\kappa}$ \cite{Nieto:2016gnp}. However, since $\rho_{m}+P_{m} \neq 0$, successful slow roll inflation requires $\rho_{m} \sim \rho_{\rm YM} \ll \rho_{\kappa}$ in order to ensure $\epsilon_{H} \ll 1$ (see eq.\ \eqref{eq:Hdot} below).

The equation of motion for the gauge-field vacuum expectation value (vev) can be derived from the action in eq.\ \eqref{eqn:minisupact}. In terms of the variable $\psi=\phi/a$, it reads
\begin{align}
\label{eq:psibackgroundeom}
\ddot \psi + {3 H \dot \psi  } + \psi \dot H
+   {2\kappa g^2 \psi^3\dot \psi^2 \over (1+\kappa g^2 \psi^4)}+ {\psi (2H^2 + 2g^2 \psi^2 +m^2 )  \over (1+\kappa g^2 \psi^4)}   = 0 .
\end{align}
Variation of the action with respect to the lapse function, $N$, and the scale factor, $a$, yield the Friedmann constraint
\begin{align}\label{eqn:Friedmann}
H^{2} =  &  \frac{1}{2}\frac{\dot\phi^2}{a^2} + \frac{1}{2}g^{2}\frac{\phi^4}{a^4}+ \frac{1}{2}m^{2}\frac{\phi^2}{a^2} + \frac{1}{2}\kappa \frac{ g^2\phi^4\dot{\phi}^2}{a^6} ,
\end{align}
and the equation of motion for the metric, 
\begin{align}\label{eq:Hdot}
\dot{H} = - \frac{\dot{\phi}^2}{ a^2} - g^2\frac{\phi^4}{a^4} - \frac{1}{2}m^2\frac{\phi^2}{a^2}.
\end{align}
We introduce the standard Hubble slow-roll parameters,
\begin{align}
\epsilon = -\frac{\dot{H}}{H^2}, \quad \eta = -\frac{\ddot{H}}{2H\dot{H}} = \epsilon - \frac{\dot{\epsilon}}{2\epsilon H},
\end{align}
as well as a parameter that characterizes the slow roll of the gauge vev
\begin{align}
\delta = -\frac{\dot{\psi}}{H\psi}.
\end{align}
To measure the various contributions to the mass of the gauge field fluctuations, we define the dimensionless mass parameters in units of the Hubble scale, 
\begin{align}
\label{eq:definegammaM}
\gamma \equiv \frac{g^2\psi^2}{H^2}, \quad M \equiv \frac{m}{H} \, .
\end{align}
The Hubble parameter can be approximated by \cite{Nieto:2016gnp, Adshead:2017hnc} 
\begin{align}
H^2 \approx   {g^2 \epsilon \over \gamma \left(1+\gamma +{M^2\over 2} \right) } \, ,
\end{align}
and  the total number of $e$-folds of inflation is determined by the integral
\begin{align}
\label{eq:Nint}
N_e = \int_{t_i}^{t_f} H \d t =  -\int_{H_i}^{H_f} \frac{\d H}{H \epsilon}\approx-\frac{1}{2\psi^2}\int_{H_i}^{H_f} \frac{\d (H^2)}{H^2 (1+ \gamma+ M^2/2)}.
\end{align}
Following \cite{Nieto:2016gnp}, we use $\theta =  \gamma+ M^2/2$ as the integration variable and evaluate the integral expressed only in terms of the initial values  
\begin{align}
\label{eq:Napprox-iso}
N_e \approx -\frac{1}{2\psi^2}\ln{\(\frac{1}{\theta}+1\)} \Bigr\vert_{\theta_i}^{\theta_f}= \frac{1}{2\psi^2}\ln\({\frac{1+ \gamma_{\rm in}+ M_{\rm in}^2/2}{\gamma_{\rm in}+M_{\rm in}^2/2}}\).
\end{align}
In the slow-roll approximation, $\psi \approx$ const., and the background solution is fully specified by three independent variables out of the original set $\{H_{_{\rm in}}, \psi, \dot\psi, \gamma, M, g, \kappa\}$. This can be seen as follows. Eqs.\ \eqref{eq:psibackgroundeom},  \eqref{eqn:Friedmann}, and \eqref{eq:definegammaM}  reduce the number of independent variables to four, while the parameter $\kappa$ can be eliminated from the evolution equations by redefining time $t \to \sqrt{\kappa}t$ and the gauge coupling $g \to g/\sqrt{\kappa}$. The background trajectories are therefore specified by choosing intial values for $\{\psi_{\rm in}, \gamma_{\rm in}, M_{\rm in}\}$, or $\{N_{\rm in}, \gamma_{\rm in}, M_{\rm in}\}$ via eq.\ \eqref{eq:Napprox-iso}. The parameter $\kappa$ is determined by matching the amplitude of the curvature perturbations to the observed value \cite{Namba:2013kia, Adshead:2017hnc}.

\section{Massive Gauge-flation in a homogeneous, anisotropic background}\label{sec-static}

In this section, we consider the effect of putting massive Gauge-flation in a homogeneous, but anisotropic background.  We begin by deriving the action and equations of motion before studying the fixed points of the system and the solutions of the equations of motion.

\subsection{Action and equations of motion}
We consider the axially symmetric Bianchi type-I metric 
\be\label{metric}
\d s^2=-N^2\d t^2+e^{2\alpha}\big(e^{-4\sigma}\d x^2+e^{2\sigma}(\d y^2+\d z^2)\big),
 \ee
and we choose the following axially symmetric ansatz for the gauge field configuration \cite{Maleknejad:2011jr, Maleknejad:2012as}, 
\begin{align} \label{field-config}
A^a_{~0} = 0,\qquad A^a_{~i}=e^a_{~i}\psi_i,
\end{align}
where
\begin{align}
e^1_{~1}= e^{\alpha-2\sigma},\qquad e^2_{~2}= e^3_{~3}= e^{\alpha+\sigma}, \qquad\psi_1\equiv\frac{\psi}{\lambda^2},\qquad \psi_2=\psi_3\equiv\lambda\psi.
\end{align}
The parameter $\lambda$ characterizes the gauge-field anisotropy, with $\lambda = 1$ corresponding to the isotropic limit. Inserting the metric, eq.\ \eqref{metric}, and the ansatz for the gauge field, eq.\ \eqref{field-config}, into the action in eq.\ \eqref{eqn:GFaction22}, we obtain the reduced action 
\begin{align}\label{reduced-action}
\mathcal{L}_{\rm red} &= \frac{a^3}{N}\Bigg[-3\dot \alpha^2+\dot\sigma^2\(3+\frac{(2+\lambda^6)}{\lambda^4}\frac{\phi^2}{a^2}\)
+\dot\sigma\frac{\big(\lambda^{-4}(\lambda^6-1)\phi^2\dot{\big)}}{a^2}+\frac{(1+2\lambda^6)}{2\lambda^4}\frac{\dot\phi^2}{a^2}\n
&\quad +2\frac{(\lambda^6-1)}{\lambda^4}\frac{\dot\lambda}{\lambda}\frac{\dot\phi\phi}{a^2}+\frac{(2+\lambda^6)}{\lambda^4}\frac{\dot\lambda^2}{\lambda^2}\frac{\phi^2}{a^2}
-N^2\frac{(2+\lambda^6)}{2\lambda^2}\frac{g^2\phi^4}{a^4}+
\frac{3}{2}\frac{\kappa g^2\phi^4}{a^4}\frac{\dot\phi^2}{a^2}\Bigg]  \n
&\quad -a^3 N\[\frac{m_1^2+(m_2^2+m_3^2)\lambda^6}{2\lambda^4}\frac{\phi^2}{a^2}\]\\
&= \mathcal{L}_{\rm GF} -a^3 N\[\frac{m_1^2+(m_2^2+m_3^2)\lambda^6}{2\lambda^4}\frac{\phi^2}{a^2}\]
, \label{massive-Lag-red}
\end{align}
where $\phi(t)    = a(t)\psi(t)= e^{\alpha(t)} \psi(t) $. Here and for the rest of this paper, we denote by $a$ the isotropic or volume scale factor $a = e^\alpha$, and by $H$ the isotropic expansion rate $H = \dot\alpha$.

Note that the reduced Lagrangian in eq.\ \eqref{reduced-action} does not depend explicitly on $\sigma$, and thus the canonically  momentum conjugate to $\sigma$ is conserved \cite{Maleknejad:2011jr}.  Upon imposing the initial condition that the anisotropy $\dot\sigma = 0 $ for isotropic gauge field, that is when $\lambda = 1$, we obtain an expression for $\dot\sigma$
\be \label{massless-sigma} 
\dot\sigma=-\frac{\frac{d}{dt}\big(\lambda^{-4}(\lambda^6-1)\phi^2{\big)}}{2a^2\big(3+\lambda^{-4}(2+\lambda^6)\frac{\phi^2}{a^2}\big)}\, .
\ee
Next, we derive the energy-momentum tensor and gravitational field equations. As before, we treat $\mathcal{L} ^{\rm YM},  \mathcal{L} ^{\kappa}$ and $\mathcal{L} ^{m}$ separately.

\paragraph{Yang-Mills pressure and energy density.}

The Yang-Mills term makes non-zero contributions to the stress tensor as
\begin{align}\nn
\qquad T^{0}{}_{0}{}^{\rm YM} 
 =&-\frac{1}{2\lambda^4}\(\frac{\dot\phi}{a}-2\(\dot\sigma+\frac{\dot\lambda}{\lambda}\)\frac{\phi}{a}\)^2
-\lambda^2\(\frac{\dot\phi}{a}+\(\dot\sigma+\frac{\dot\lambda}{\lambda}\)\frac{\phi}{a}\)^2
-\frac{(2+\lambda^6)}{2\lambda^2}\frac{g^2\phi^4}{a^4},\\\nn
 T^1{}_{1}{}^{\rm YM}  =& -\frac{1}{2\lambda^4}\(\frac{\dot\phi}{a}-2\(\dot\sigma+\frac{\dot\lambda}{\lambda}\)\frac{\phi}{a}\)^2
+\lambda^2\(\frac{\dot\phi}{a}+\(\dot\sigma+\frac{\dot\lambda}{\lambda}\)\frac{\phi}{a}\)^2
+ \frac{(2-\lambda^6)}{2\lambda^2}\frac{g^2\phi^4}{a^4},\\\nn
 T^{2}{}_{2}{}^{\rm YM} = &  
 \frac{1}{2\lambda^4}\(\frac{\dot\phi}{a}-2\(\dot\sigma+\frac{\dot\lambda}{\lambda}\)\frac{\phi}{a}\)^2
+ \frac{\lambda^4}{2}\frac{g^2\phi^4}{a^4},\\
T^{3}{}_{3}{}^{\rm YM} = & T^{2}{}_{2}{}^{\rm YM}.
\end{align}
We  extract the energy density and rewrite the pressures in terms of isotropic $(P^{_{\rm YM}})$ and anisotropic parts ($\tilde{P}^{_{\rm YM}}$) as
\begin{align}
P_{_{\rm YM}}= & \frac13 \rho_{_{\rm YM}},\qquad
P^{_{\rm YM}}_{1} = P_{_{\rm YM}}  - \frac{2}{3} \tilde{P}_{_{\rm YM}},\qquad
P^{_{\rm YM}}_{2} =  P^{_{\rm YM}}_{3} = P_{_{\rm YM}}+ \frac{1}{3} \tilde{P}_{_{\rm YM}},
\end{align}
where
\begin{align} \label{rho_YM} \nn
 \rho_{_{\rm YM}} = & \frac{1}{2\lambda^4}\(\frac{\dot\phi}{a}-2\(\dot\sigma+\frac{\dot\lambda}{\lambda}\)\frac{\phi}{a}\)^2
+\lambda^2\(\frac{\dot\phi}{a}+\(\dot\sigma+\frac{\dot\lambda}{\lambda}\)\frac{\phi}{a}\)^2
+\frac{(2+\lambda^6)}{2\lambda^2}\frac{g^2\phi^4}{a^4},\\
\tilde{P}_{_{\rm YM}}= & \frac{1}{\lambda^4}\(\frac{\dot\phi}{a}-2\(\dot\sigma+\frac{\dot\lambda}{\lambda}\)\frac{\phi}{a}\)^2
-\lambda^2\(\frac{\dot\phi}{a}+\(\dot\sigma+\frac{\dot\lambda}{\lambda}\)\frac{\phi}{a}\)^2
- \frac{(1-\lambda^6)}{\lambda^2}\frac{g^2\phi^4}{a^4}.
\end{align}

\paragraph{$\kappa$-term pressure and energy density.}
The contribution to the stress tensor from the $\kappa$-term is the same as in an isotropic background, 
\be
T_{\mu\nu}^{\kappa} = - \frac{3}{2}\frac{\kappa g^2\phi^4 \dot\phi^2}{a^6N^2} g_{\mu\nu}.
\ee
We then identify the energy density and (isotropic) pressure associated with this term as
\be \label{rho_kappa}
\rho_{\kappa} = \frac{3}{2}\frac{\kappa g^2\phi^4 \dot\phi^2}{a^6}, \quad 
P_{\kappa} = - \frac{3}{2}\frac{\kappa g^2\phi^4 \dot\phi^2}{a^6}.
\ee

\paragraph{Symmetry-breaking pressure and energy density.}
The stress tensor due to the symmetry breaking terms is obtained in a straightforward fashion as
\begin{align}
T_{\mu\nu}^{m} & = m_a^{~2} A^{a}_{~\mu} A^{a}_{~\nu} - \frac{g_{\mu\nu}}{2} m^{~2}_aA_{\sigma}^{~a} A^{a\,\sigma}.
\end{align}
We then extract the energy densities and pressures due to the gauge field mass terms
\be \label{rho_massive}
\rho_{m} =  \ddfrac{m_1^2+(m_2^2+m_3^2)\lambda^6}{2\lambda^4}  \ddfrac{\phi^2}{a^2},\\
\ee
and
\begin{align}  \label{p1_massive}
P^{m}_{1} =& \ddfrac{m_1^2-(m_2^2+m_3^2)\lambda^6}{2\lambda^4}  \ddfrac{\phi^2}{a^2}, \quad
P^{m}_{2} =  \ddfrac{-m_1^2+(m_2^2-m_3^2)\lambda^6}{2\lambda^4} \ddfrac{\phi^2}{a^2},\\\nn
%f
P^{m}_{3} = &  \ddfrac{-m_1^2+(m_3^2-m_2^2)\lambda^6}{2\lambda^4} \ddfrac{\phi^2}{a^2}.
\end{align}
The self-consistency of our axially symmetric ansatz requires that we take $m_2=m_3$. We again decompose the pressure into isotropic and anisotropic parts as
\begin{align}
P^m_{1} = P_m - \frac{2}{3} \tilde{P}_m,\quad
P^m_{2} =  P^m_{3} = P_m + \frac{1}{3} \tilde{P}_m,
\end{align}
where
\begin{align}
\rho_{m}= \ddfrac{m_1^2}{2\lambda^4}+ m_2^2\lambda^2 , \quad \tilde{P}_m = -\ddfrac{ m_1^2 }{\lambda^4}+ m_2^2\lambda^2, \quad P_m = -\frac{1}{3}\rho_{m}.
\end{align}

\paragraph{Total pressure and energy density and gravitational field equations.}
In sum, we find the total contributions to the energy density and the isotropic and anisotropic pressures
\begin{align}
\rho =  \rho_{\kappa} +   \rho_{_{\rm YM}}+\rho_m ,\quad P = &  -\rho_\kappa + \frac{1}{3}\rho_{_{\rm YM}} - \frac{1}{3}\rho_m, \quad \tilde{P} = \tilde P_{_{\rm YM}}+\tilde P_{m},\\
P_1 =  P  -\frac23\tilde P, & \quad   P_2 = P +\frac13 \tilde P.
\end{align}

Variation of the action with respect to $N$, $\alpha$, and $\sigma$ yields the gravitational field equations,
 \begin{align}
\dot{\alpha}^2-\dot{\sigma}^2= & \frac{\rho}{3}, \label{dot-alpha^2}\\
\ddot{\sigma}+3\dot{\alpha}\dot{\sigma}=& \frac{P_2-P_1}{3}=\frac{\tilde P}{3}, \label{ddot-sigma}\\
\ddot{\alpha}+3\dot{\sigma}^2=& -\frac{3\rho+P_1+2P_2}{6}=-\frac23\rho_{_{\rm YM}}- \ddfrac13\rho_{_{m}} \label{ddot-alpha}.
\end{align}
We note that,  $\tilde{P}$ vanishes in the isotropic limit, i.e $\lambda ^2 = 1$, only if $m_1=m_2$. Since $ \tilde{P} $ sources the anisotropy in the spatial slicing, $\dot\sigma$ (from eq.\ \eqref{ddot-sigma}), in the case that $m_1 \neq m_2$, an initially isotropic solution can be driven toward an anisotropic solution.

We are primarily interested in inflation and so we first determine the conditions under which accelerated expansion is possible.  Through eqs.\ \eqref{dot-alpha^2} and \eqref{ddot-alpha}, we find the expression for $\epsilon = - \dot{H}/H^2 = - \ddot\alpha/\dot\alpha^2$, 
\be\label{eqn:epsilon_static}
\epsilon=\ddfrac{2\rho_{_{\rm YM}}+\rho_{m}+9\dot\sigma^2}{\rho_\kappa+\rho_{_{\rm YM}}+\rho_{m}+3\dot\sigma^2}.
\ee
For inflation to proceed we require $\epsilon < 1$, which from eq.\ \eqref{eqn:epsilon_static} amounts to the requirement  $\rho_{\kappa}\gg\rho_{_{\rm YM}}, \rho_{m},\dot\sigma^2$ for sufficiently long time.  In the remainder of this section, we study the fixed points of the motion, the effect of the anisotropy on the inflationary solutions as well as the evolution of the anisotropy of the spatial slices.

%%%%%%%%%%%%%%%%%%%%%%%%%%%%%%%%%%%%%%%%%%%%%%%%%%%
\begin{figure}[t!]
\includegraphics[width=\textwidth]{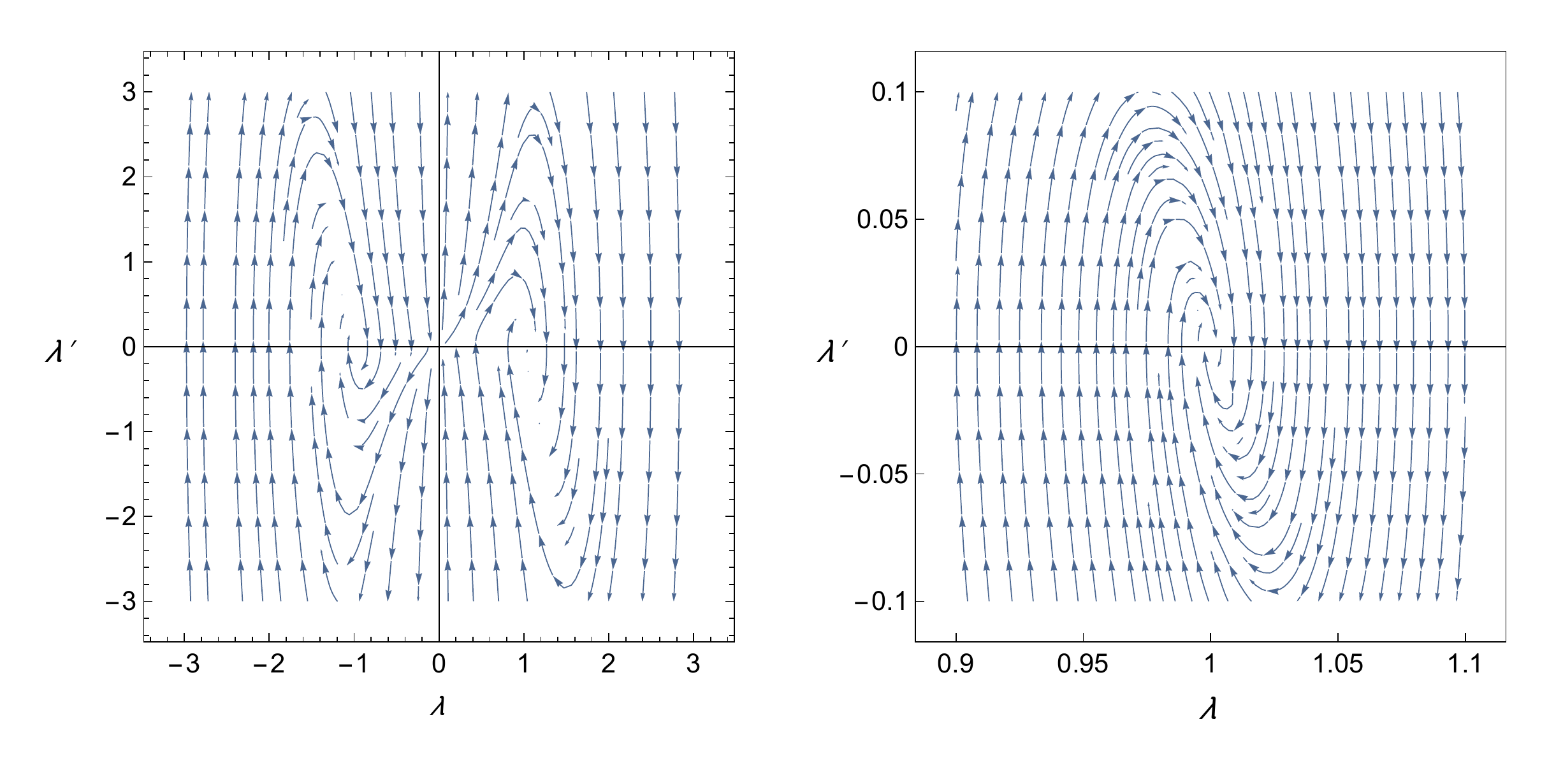}
\caption{Phase portrait of the gauge-field anisotropy parameter, $\lambda$, when $M_1=M_2$ and $\gamma = 3$. In the left panel, we exhibit the symmetry of the solutions under the reflection $\lambda \to -\lambda$, ${\lambda}' \to -\lambda'$. In the right panel, we display the behavior near the (isotropic) fixed point at $\lambda = 1$.} \label{fig:phaseplanem0}
\end{figure}

\begin{figure}[t!]
\includegraphics[width=\textwidth]{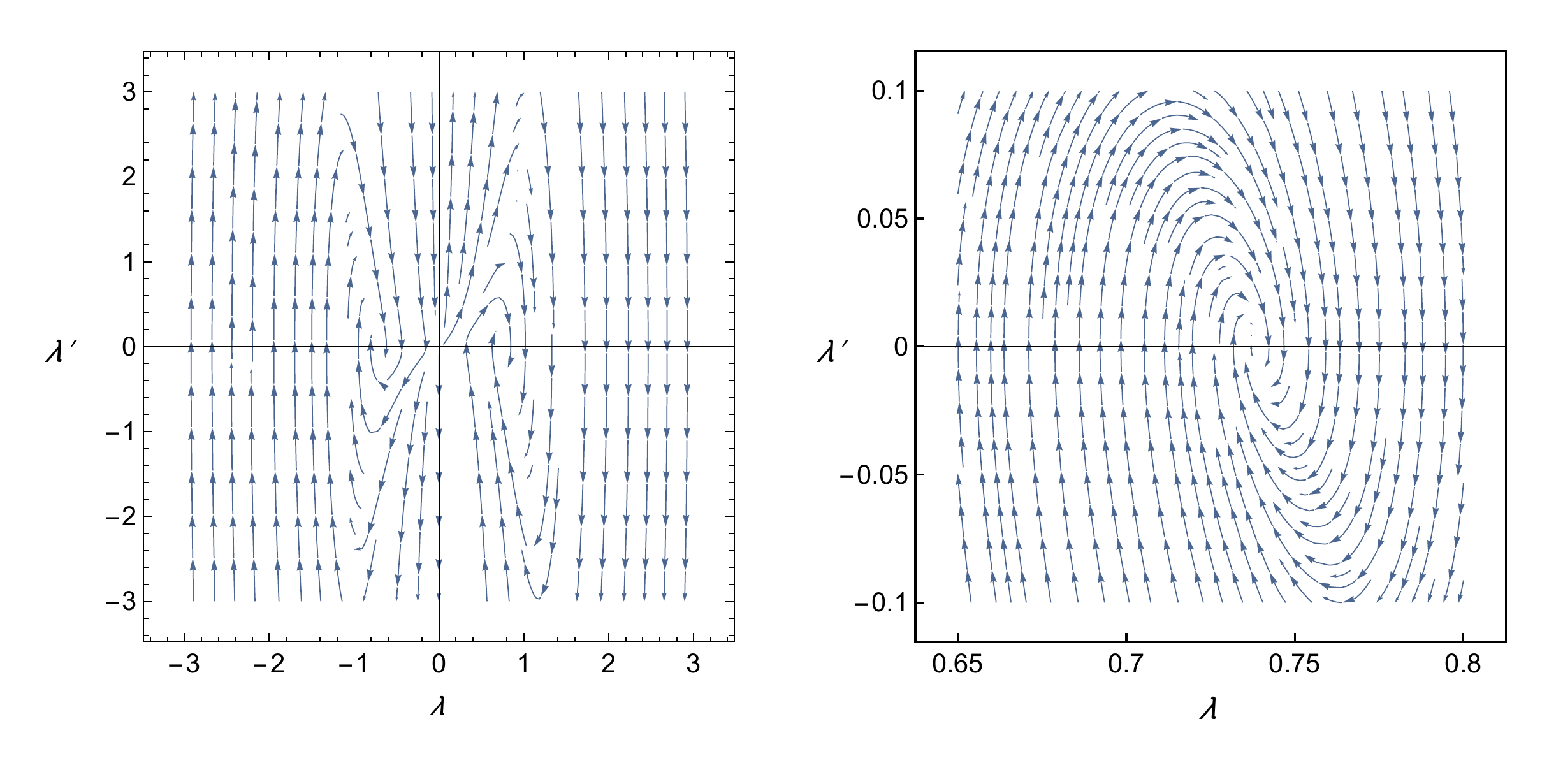}
\caption{Phase portrait of $\lambda$ in an example with an anistropic fixed point. Here $M_2 = 5 M_1 = 5$ and $\gamma = 3$. In the left panel, we show that the symmetry under the reflection $\lambda \to -\lambda$, ${\lambda}' \to -\lambda'$ is preserved. In the right panel, we display the details of the anisotropic fixed point which occurs at  $\lambda \sim 0.74$.} \label{fig:phaseplanem1}
\end{figure}

%
%%%%%%%%
\subsection{Phase-plane and fixed points of the motion}
%%%%%%%
%

The non-linear nature of the equations for the gauge-field anisotropy means that numerical analysis is required to completely characterize the trajectories. However, we can gain some understanding of the dynamics by studying the phase plane and identifying the (approximate) fixed points of the motion.

Varying the action with respect to $\lambda$, we find the equation of motion
\begin{align}\nn
\(\(2+\lambda^6\)\(\lambda\ddot\lambda+\dot\alpha\lambda\dot\lambda+2\frac{\dot\phi}{\phi}\lambda\dot\lambda\)-6\dot\lambda^2\)
\frac{\phi^2}{a^2} & +\lambda^2\(\lambda^6-1\)\(\frac{\phi\ddot\phi}{a^2}+\frac{\dot\alpha\phi\dot\phi}{a^2}+\frac{\lambda^2g^2\phi^4}{a^4}\)\\ &  -\(m_1^2 - m_2^2 \lambda^6\) \lambda^2 \frac{\phi^2}{a^2}=0,
\end{align}
Working in the slow-roll limit --- treating $\psi \approx$ const.\ and  $\dot\phi \approx \dot\alpha \phi$ --- we obtain
\be
(2+\lambda^6)(\lambda\ddot\lambda+3\dot\alpha\lambda\dot\lambda)-6\dot\lambda^2
+\lambda^2(\lambda^6-1)(2+\lambda^2\gamma)\dot\alpha^2 -\lambda^2 (m_1^2 - m_2^2 \lambda^6) \simeq0 \, . \label{eq: lambda-evo}
\ee
We replace the comic time derivative with conformal time derivative, denoted here by a prime, so that the Hubble parameter, $H = \dot\alpha$, does not show up explicitly, 
\be
(2+\lambda^6)(\lambda\lambda''+3\lambda\lambda')-6\lambda'^2
+\lambda^2(\lambda^6-1)(2+\lambda^2\gamma) -\lambda^2 (M_1^2 - M_2^2 \lambda^6) \simeq0, \label{eq: lambda-evo-conf}
\ee
where we have defined, 
\bea
M_1= \frac{m_1}{H},\quad  M_2= \frac{m_2}{H}  \,  .                                                                
\eea

We look for fixed points of the motion\footnote{These fixed points will of course vary slowly with the evolution of the background, however, it is easy to see that the timescale associated with the evolution of the fixed point is much slower than the timescale associated with the evolution of $\lambda$ near to the fixed point.} by setting ${\lambda}'' = {\lambda}' = 0$. We first observe that the phase plane is invariant under replacing $\{\lambda, {\lambda}'\} \to \{-\lambda, -{\lambda}'\}$. Next, we note that when the gauge field masses either vanish, or are equal $M_1^2 = M_2^2$, we recover the known result that the only fixed point is the isotropic fixed point at $\lambda = \pm 1$ \cite{Maleknejad:2011jr}. The phase plane for the massless case is displayed in figure \ref{fig:phaseplanem0}.  When the gauge field masses are unequal, $M_1\neq M_2$, the fixed points are driven away from the isotropic fixed point at $\{\lambda, \lambda' \}= \{\pm 1,0\}$. In figure \ref{fig:phaseplanem1} we show the effect of unequal gauge field masses on the phase plane. In the example shown here, $M_2 = 5 M_1$ and the fixed point occurs at $\lambda \sim 0.74$ and remains an attractor. Note that the nature of the anisotropy ($\lambda > 0$ and prolate, or $\lambda < 1$ and oblate) depends on the relative sizes of $M_1$ and $M_2$.

\begin{figure}
\includegraphics[width = 1.05\textwidth]{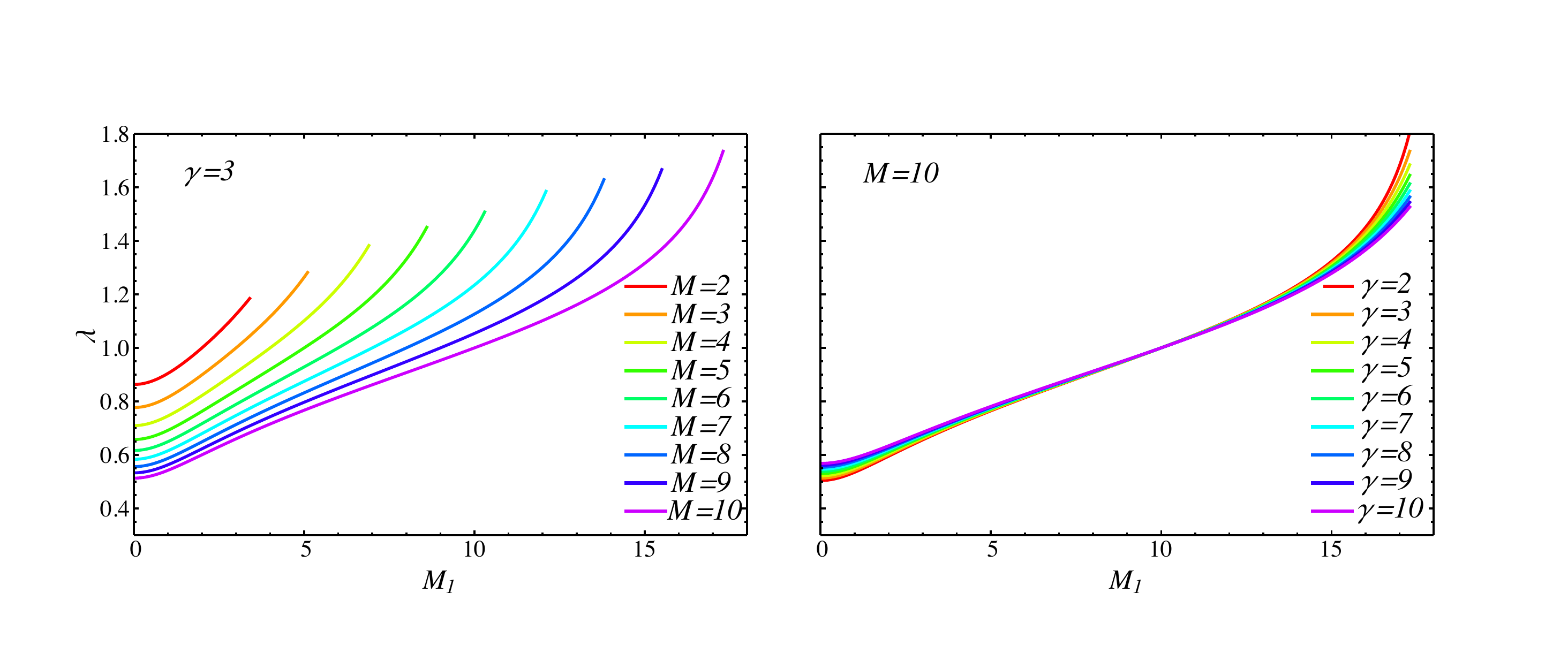}
\caption{Fixed points of the gauge field anisotropy parameter, $\lambda$, as a function of $M_1$. Each curve is generated by fixing ${M}$ and $\gamma$ and solving eq.\ \eqref{eq:stable-lambda} while varying $M_1$. The endpoints of the curve correspond to the extrema of the allowed values of $M_1$ at fixed $M = \sqrt{(M_1^2+2M_2^2)/3}$ --- recall that $M_2\geq 0$. Left panel: We fix $\gamma = 3$, and allow ${M}$ to range from 2 to 10 (from red to purple). Right panel: We fix ${M}=10$, while $\gamma$ ranges from 2 to 10 (from red to purple).}\label{fig: lambda-constraint}
\end{figure}

It is possible to find the fixed points analytically. Returning to eq.\ (\ref{eq: lambda-evo}) and setting $\ddot{\lambda} = \dot{\lambda} = 0$, we find
\be
\label{eq:stable-lambda}
(\lambda^6-1)(2+\lambda^2\gamma) -(M_1^2 - M_2^2 \lambda^6) =0. 
\ee
Solutions of eq.\ \eqref{eq:stable-lambda} are readily found, however, since this requires solving a quartic equation (for $\lambda^2$), the resulting expressions are not particularly illuminating. We  instead show the result graphically.  To display the result on a 2D plot, we fix the average gauge field mass, $M$, and the background gauge field $\gamma$, and plot $\lambda$ as a function of one of the gauge field masses, $M_1$. The results are shown in figure \ref{fig: lambda-constraint}. Note that when $M_1 = M_2$, $\lambda = 1$, while $M_1 < M_2$  and $M_1 > M_2$ correspond to $ \lambda < 1$ and $\lambda > 1$ respectively. Varying $\gamma$ has only a small effect on the anisotropy, as we demonstrate in the right-hand panel of figure \ref{fig: lambda-constraint}.

%%%%%%%%%%%%%%%%%%
\subsection{Gauge-field anisotropy and the duration of inflation}
%%%%%%%%%%%%%%%%%%

In order to determine the effect of the anisotropy on the length of inflation, we seek to develop an approximation for the number of $e$-folds, the anisotropic version of eq.\ \eqref{eq:Nint}. The existence of the additional degree of freedom, $\lambda$, appears to make this a little more complicated. However, as we now demonstrate, to leading order in the slow-roll approximation, the evolution of $\lambda$ can be neglected.

The rate of change of $\lambda$, characterized by 
\begin{align}
\Delta\equiv  \frac{\dot\lambda}{\lambda H}, 
\end{align}
can be deduced by differentiating eq.\ (\ref{eq:stable-lambda}) 
\be
\begin{split} \label{eq:lambda-dot}
\Delta
&\approx -\frac{ 2 \left(\lambda  ^6-1\right)\left(\lambda  ^6+2\right) \left(\gamma  \lambda  ^2-\gamma  \lambda  +2\right)}{\gamma  \lambda  \left[\left(6 \lambda ^7-7 \lambda ^6-6 \lambda -13\right) \lambda ^6+2\right]-6 \lambda ^6 \left(M^2+6\right)}\,\epsilon\, ,
\end{split}
\ee
where we have defined, 
\be
M \equiv \sqrt{\frac{M_1^2+2M_2^2}{3}}\,.
\ee

\begin{figure}
\begin{center}
\includegraphics[width=0.75 \textwidth]{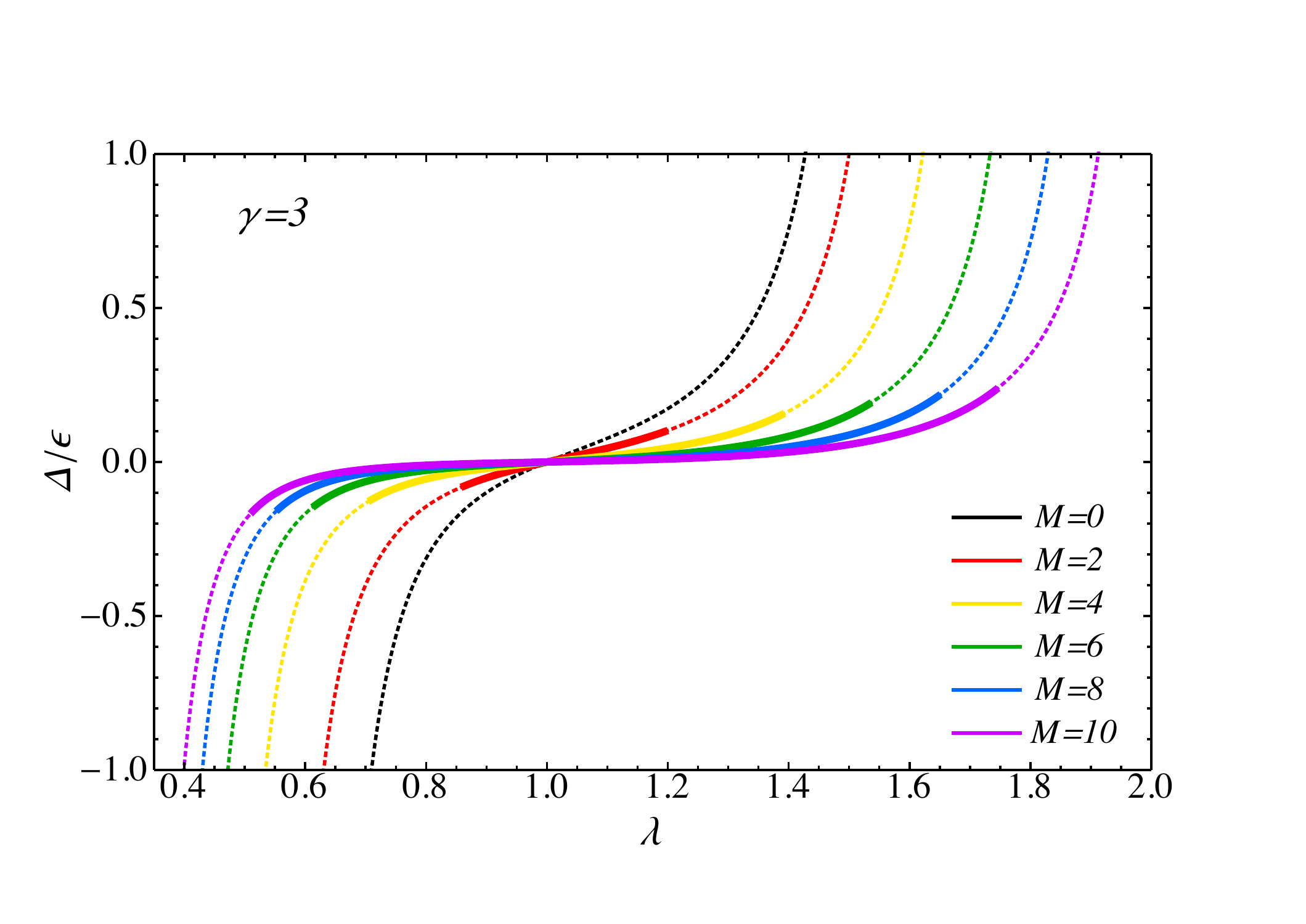}
\caption{The evolution of the field anisotropy parameter, $\lambda$, quantified by $\Delta/\epsilon$. We fix $\gamma = 3$, and ${M}$ ranges from 0 (black curve) to 10 (purple curve) in steps of $\scriptstyle{\Delta}$$M = 2$. The thinner, dotted segments of each curve implies that $M_1$ or $M_2$ is imaginary at that point. It is clear that in all desired situations, $|\Delta/\epsilon|<1$.}\label{fig: lambdot-M-2}
\end{center}
\end{figure}

\begin{figure}[h!]
\begin{center}
\includegraphics[width = 0.75\textwidth]{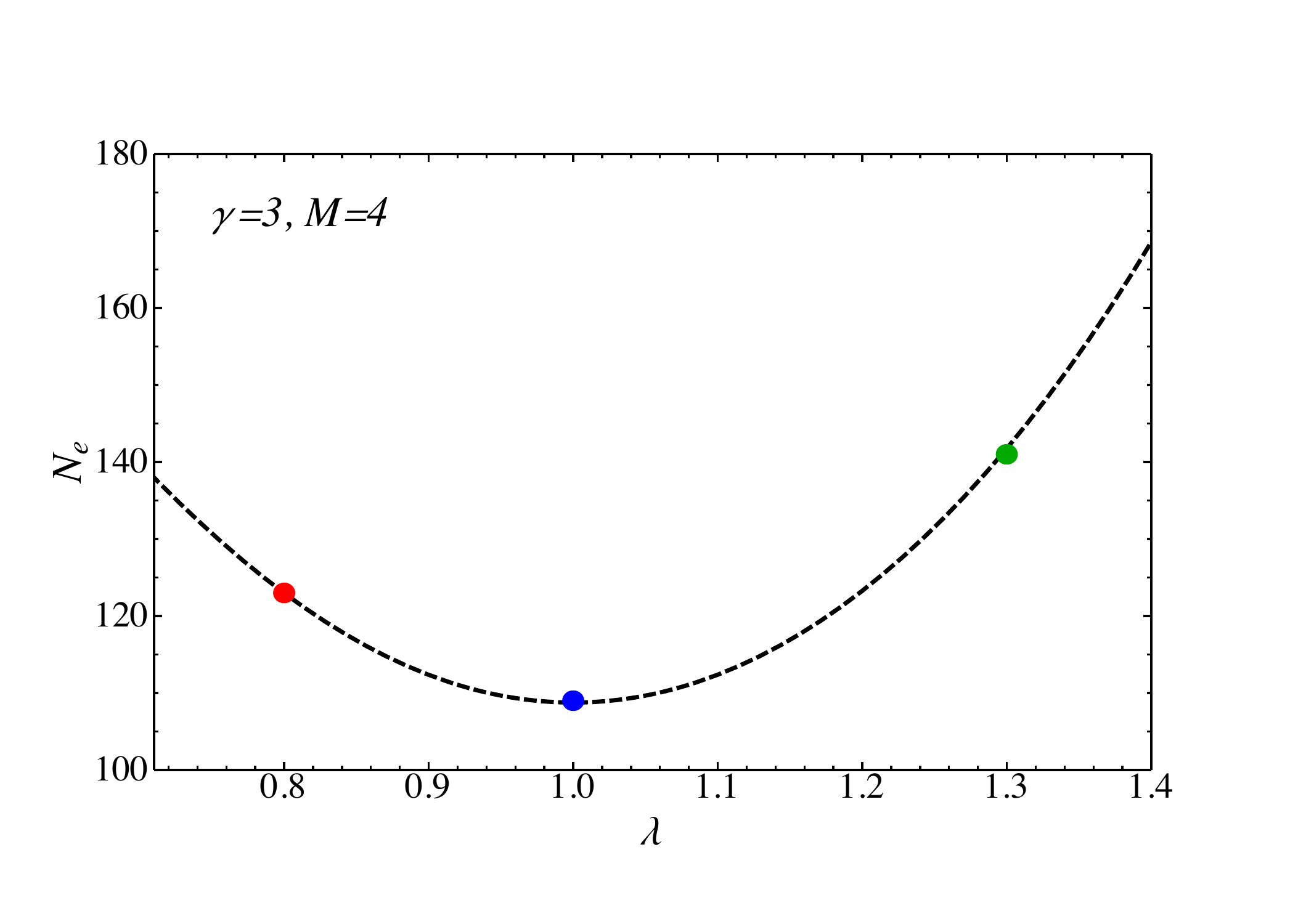}
\caption{Agreement between approximated number of $e$-folds, $N_e$, and the numerically computed results. The length of inflation, $N_e$, computed using the approximation eq.\ (\ref{eq:Napprox-aniso2}) is shown with a dashed line. The three coloured points are the values of $N_e$ calculated numerically by solving the full equations of motion. To generate the figure, we  fix the parameters $\{\psi_{{\rm in}}=0.02, \gamma_{{\rm in}} = 3, { M}_{{\rm in}} =4 \}$ while varying $\lambda_{\rm in}$.}\label{fig: 3eg-efolds}
\end{center}
\end{figure}

Figure \ref{fig: lambdot-M-2} shows $\Delta/\epsilon $ as a function of $\lambda$. In this figure, we plot eq.\ \eqref{eq:lambda-dot} as a function of $\lambda$ while fixing $\gamma$ and $M$. Since not all values of $\lambda$ are permitted at fixed $\gamma$ and $M$, we indicate the physical solutions by the thick solid curves. These thick solid lines become thin dashed curves when $M_1^2$ or $M_2^2$ are negative (according to eq.\ \eqref{eq:stable-lambda}), corresponding to unphysical values of $\lambda$. Clearly $|\Delta/\epsilon|$ is always less than unity, and in most cases is $\ll 1$.  Therefore, to a very good approximation,  $\lambda$ is approximately constant during inflation. 
%%%%%%%%
%%%%%%%%
As in section \ref{sec:backgroundeqns}, the length of the inflationary phase can be found by computing the integral 
\begin{align}
\label{eq:Nint2}
N_e = \int_{t_i}^{t_f} H \d t =  -\int_{H_i}^{H_f} \frac{\d H}{H \epsilon}\, .
\end{align}
After approximating eq.\ \eqref{eqn:epsilon_static} as
\begin{align}\label{eq:aproxepsilon}
\epsilon&
\approx \frac{\psi^2}{3} \[\(\frac{1}{\lambda^4}+2\lambda^2\)+ \(\frac{1}{2\lambda^4}M_1^2+\lambda^2M_2^2\) + \(\frac{2}{\lambda^2}+\lambda^4\) \gamma \],
\end{align}
which is compared with the numerical results in figure \ref{fig: 3eg}, we can compute the integral in eq.\ \eqref{eq:Nint2} to obtain
\begin{align}
\label{eq:Napprox-aniso2}
N_e &\approx \frac{1}{2\psi_{\rm in} ^2 } \frac{3 \lambda_{_{\rm in}}^4}{\(1+2 \lambda_{_{\rm in}}^6\)} \ln\[1+\frac{2+ \lambda_{_{\rm in}}^6}{3}\cdot {\(\frac{ 3\lambda_{_{\rm in}}^6}{1+2 \lambda_{_{\rm in}}^6} \cdot\frac12{ M}_{_{\rm in}}^{2} +\lambda_{_{\rm in}}^2\gamma_{_{\rm in}}  -\frac23 \frac{ (1-\lambda_{_{\rm in}}^6)^2}{ 1+2\lambda_{_{\rm in}}^6}\)}^{-1}\].
\end{align}
In figure \ref{fig: 3eg-efolds} we demonstrate the accuracy of our approximation to $N_e$.  We show the comparison between the length of inflation ($e$-fold number $N_e$), obtained from direct numerical calculations, and the approximation in eq.\ \eqref{eq:Napprox-aniso2}. Note that the agreement is excellent.
%%%
%%%

\begin{figure}
\includegraphics[width = \textwidth]{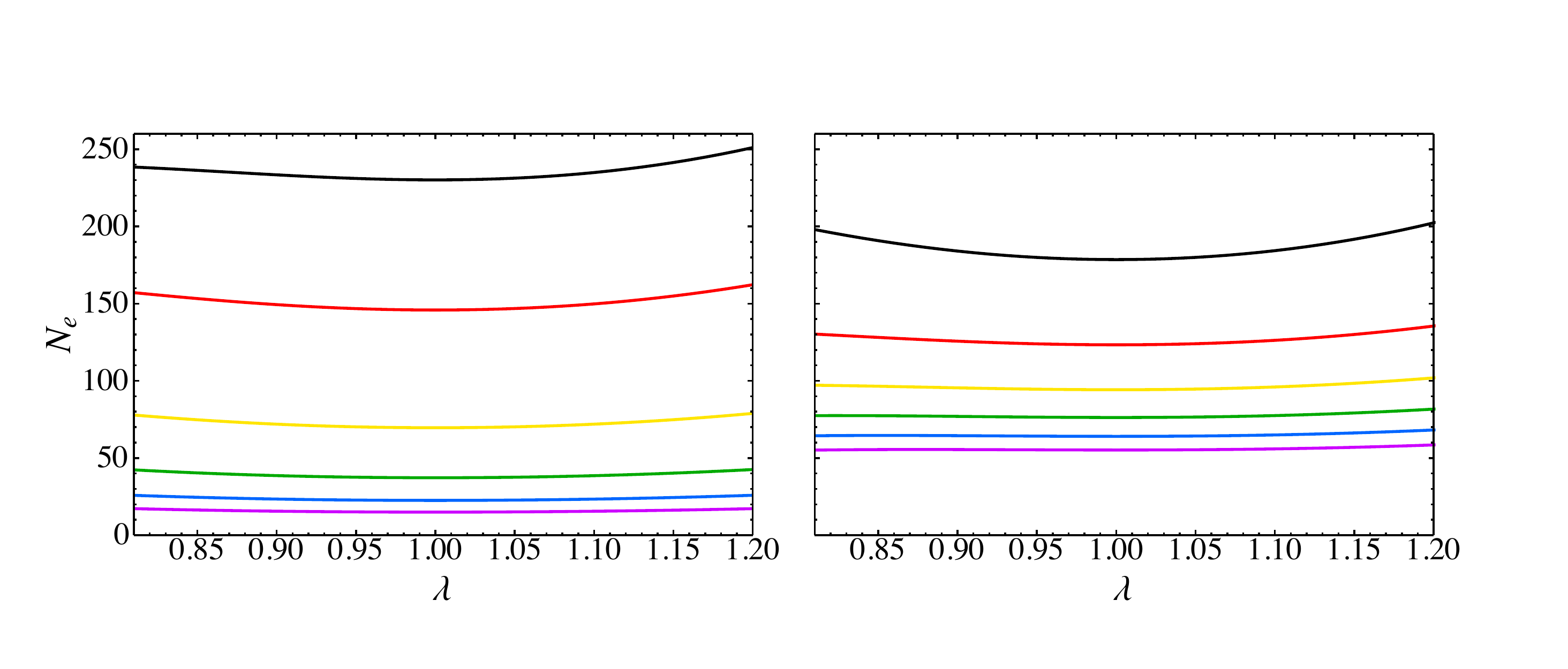}
\caption{Left panel: the effect of varying $\lambda_{\rm in}$ and $M_{\rm in}$ on $N_e$, the number of $e$-foldings of anisotropic inflation (with $\gamma_{\rm in}=3$, and $\psi_{\rm in}=0.025$). The initial value $M_{\rm in}$ varies from $0$ (top, black curve) to $10$ (lowest, purple curve) in steps of $\scriptstyle{\Delta}$$M = 2$. Right panel: the effect of varying $\lambda_{\rm in}$ and $\gamma_{\rm in}$  on the number of $e$-foldings of inflation (with $M_{\rm in}=4$, and $\psi_{\rm in}=0.025$). The initial value $\gamma_{\rm in}$ varies between 2 (top, black curve) and 12 (lowest, purple curve) in steps of $\scriptstyle{\Delta}$$\gamma = 2$.}\label{fig: efold-lambda}
\end{figure}

In figure \ref{fig: efold-lambda}, we show effect of  different parameter combinations $\{ \gamma,  M, \lambda\}$ on the length of inflation. Increasing $M$ generally shortens the inflationary period, while anisotropy ($\lambda\neq1$) generally prolongs it. That is, for fixed $M$, the length of the inflationary phase is \emph{minimized} at the isotropic fixed point at $\lambda = 1$.

As in the isotropic case, the length of inflation (on the attractor solution) is  determined by specifying the initial values of $\psi_{\rm in}$, $\gamma_{{\rm in}}$, and ${M}_{{\rm in}}$. However, in the anisotropic case we need to additionally specify the initial anisotropy, $\lambda_{\rm in}$. Though the set $\{\psi_{\rm in}, \gamma_{{\rm in}}, M_{1_{\rm in}}, M_{2_{\rm in}}\}$ is the most intuitive choice of parameterization, we choose the equivalent parametrization of $\{\psi_{\rm in}, \gamma_{{\rm in}}, { M}_{{\rm in}},\lambda_{\rm{in}}\}$ in following discussion so that we can obtain an analytic approximation for $N_e$, that is, eq.\ \eqref{eq:Napprox-aniso2}. Eq.\ \eqref{eq:Napprox-aniso2} also implies that we can instead choose to specify the set $\{N_e, \gamma_{{\rm in}}, { M}_{{\rm in}},\lambda_{\rm{in}}\} $. 

%%%%%%%%%%%%%%%%%%%%%%%%%%%%%%%%%%%%%%%%%%%%%%%%%%%

\subsubsection{Anisotropy and numerical analysis} 

\begin{figure}[h!]
\centering
\includegraphics[width=\textwidth]{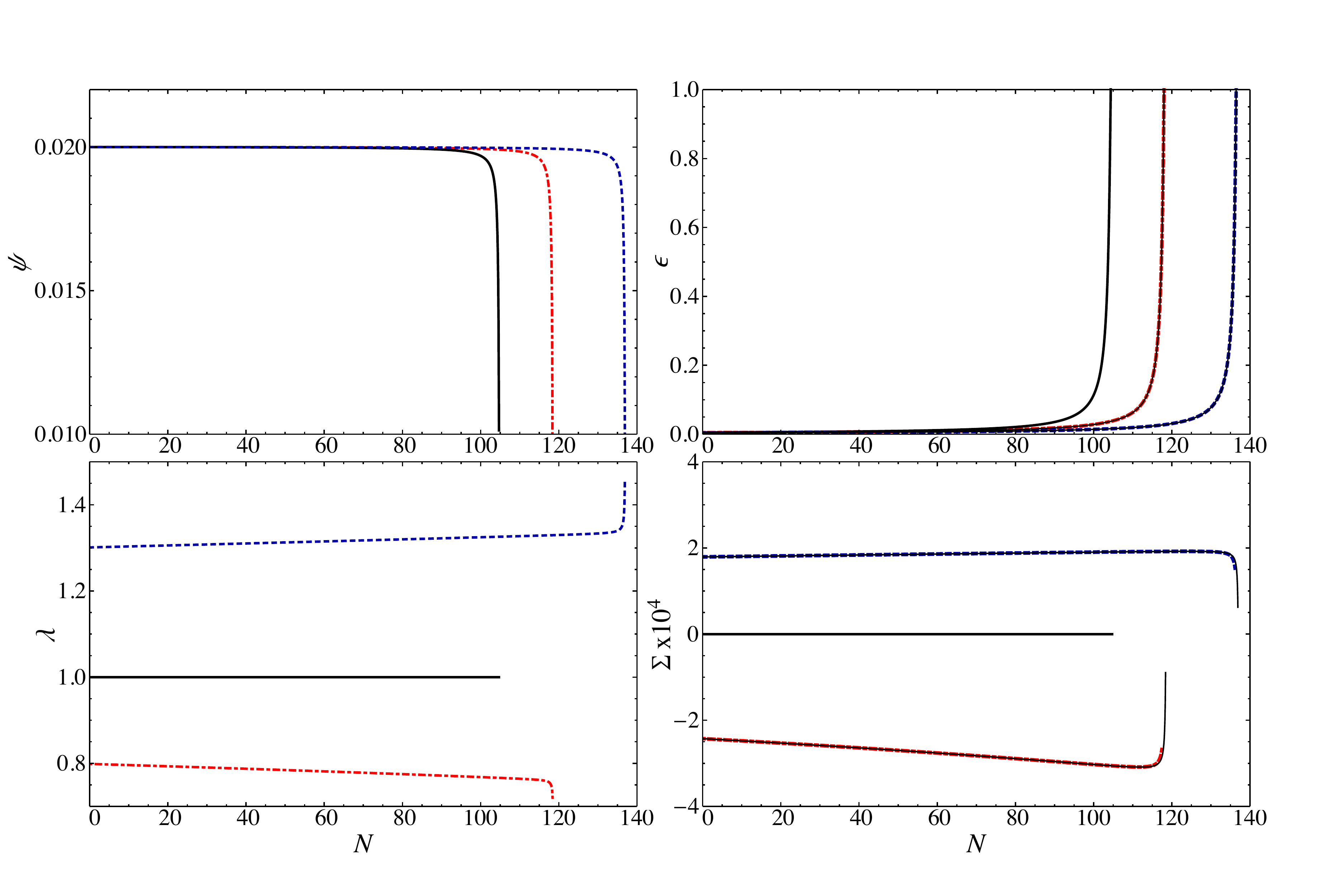}
\caption{The full numerical evolution of $\psi(N)$ (top left panel), $\epsilon = -\dot{H}/H^2$ (top right panel), $\lambda(N)$ (bottom left panel), and $\Sigma = \dot{\sigma}/H$ (bottom right panel). To generate the curves, the initial values of the parameters were chosen to be  $\psi_{\rm in} = 0.02$, $\gamma_{\rm in}=3$, $M_{\rm in }=4$, while $\lambda_{\rm in}$ is taken to be $\lambda_{\rm in} = 0.8$ (red dot-dashed curves), $\lambda_{\rm in} = 1$ (black, solid curves), and $\lambda_{\rm in} = 1.3$ (dotted blue curves). The thin black curves show the approximations for $\epsilon$ (top right panel) corresponding to eq.\ \eqref{eq:aproxepsilon} and  the approximation to $\Sigma$ (bottom right panel) eq.\ \eqref{eq:sigma-app}.}\label{fig: 3eg}
\end{figure}
%%%%%%%%%%%%%%%%%%%%%%%%%%%%%%%%%%%%%%%%%%%%%%%%%%%

%%%%%%%%%%%%%%%%%%%%%%%%%%%%%%%%%%%%%%%%%%%%%%%%%%%

We now examine the anisotropy of the spatial slices and study the full evolution  of the system by solving the system of equations numerically. The anisotropy of the spatial slices is characterized by the quantity
\begin{align}
\Sigma \equiv \frac{\dot{\sigma}}{H}.
\end{align}
Theoretically, during slow-roll inflation, $\Sigma$ is bounded by $\sqrt{\epsilon}$, where $\epsilon = -\dot{H}/H^2$ is the isotropic slow-roll parameter \cite{Maleknejad:2012as}. Specifically, 
\begin{align}
\frac{\Sigma}{\sqrt{\epsilon}} < \frac{1}{\sqrt{3}}.
\end{align}
For our model, we can derive an approximation to $\Sigma$ using eq.\ \eqref{massless-sigma}. In the slow-roll limit, and on the anisotropic attractor solution we find 
\be
\begin{split}
\Sigma &= \ddfrac{\(\lambda^2+\frac{2}{\lambda^4}\)\Delta+\(\lambda^2-\frac{1}{\lambda^4}\)(1-\delta)}{\frac{3}{\psi^2}+\lambda^2+\frac{2}{\lambda^4}}\approx \frac{\psi^2}{3}\(\lambda^2-\frac{1}{\lambda^4}\) . \label{eq:sigma-app}
\end{split}
\ee
We compare this approximation to the numerical results in figure \ref{fig: 3eg}. Eq.\ \eqref{eq:sigma-app} remains an excellent approximation to the full evolution until near the end of inflation where the slow-roll approximation begins to fail. As expected, the anisotropy vanishes in the limit $\lambda \to 1$. Further, the sign of the effect depends on the geometry---whether $\lambda > 1$ or $\lambda < 1$. Comparing eq.\ \eqref{eq:sigma-app} to eq.\ \eqref{eq:aproxepsilon} it is apparent that
\begin{align}
\Sigma & \sim \mathcal{O}(\epsilon).
\end{align}
Therefore, in the slow-roll regime we expect that we cannot approach this bound, unless we are near to the end of inflation where $\epsilon\lesssim 1$.  Anisotropy of roughly same magnitude was discovered in  power-law $k$-inflation \cite{Ohashi:2013pca,Do:2017onf}.

We now consider the numerical evolution of the full system. Figure \ref{fig: 3eg} shows the evolution of $\{\psi, \lambda, \epsilon, \Sigma\}$ in three cases corresponding to $\lambda > 1$, $\lambda = 1$ and $\lambda < 1$. The case where $\lambda_{\rm in} =1$ is equivalent to setting $m_1=m_2=m_3$, and when $m_i = 0$ we recover the results from \cite{Maleknejad:2011jr}. The cases where $\lambda_{\rm in} =0.8$ and $1.3$  correspond to the case when the spacetime is elongated along the axial and radial direction respectively. The shapes of the trajectories of $\epsilon$ and $\psi$ are similar to those in the isotropic scenario, and $\lambda$ and $\Sigma$ deviate from their isotropic fix points as expected from eqs.\ \eqref{eq:stable-lambda} and \eqref{eq:sigma-app} respectively. In both instances, the inflationary epoch is prolonged compared to the isotropic case.

%%%%%%%%%%%%%%%%%%%%%%%%%%%%%%%%%%%%%%%%%%%%%%%%%%%%%%%%%%%%%%%%%%%%%%%%%%%
\section{Dynamical symmetry breaking during inflation}\label{sec:dynamical}
%%%%%%%%%%%%%%%%%%%%%%%%%%%%%%%%%%%%%%%%%%%%%%%%%%%%%%%%%%%%%%%%%%%%%%%%%%%

In this section, we consider the situation where the symmetry is broken dynamically during inflation. That is, we allow the vev of $\Psi$ to evolve on its potential $ V(\Psi)$, and consider the full action at eq.\ \eqref{eqn:HiggsGF}.
We assume that $\Psi$ depends only on time and we  take $\Psi$ to transform under the vector representation of SU(2).  Furthermore, since the rest of the Lagrangian is invariant under SU(2) rotations, we can perform a rotation to make the vector to point along any particular direction in the internal space. We therefore choose $\Psi$ in the configuration
\begin{align}
\Psi(t) =\left( \begin{matrix} \zeta(t) \\ 0 \\ 0 \end{matrix} \right),
\end{align}
in which case the gauge field mass matrix (due to the scalar) reads
 \be
m^2_{ab} =g^2 \left( \begin{matrix} ~0  & 0 & 0   \\  ~0 & ~\tiny{\zeta^2} & ~0   \\  ~0 & 0 &  \tiny{\zeta^2}  \end{matrix} \right).
\ee
Inserting this form  into the action in eq.\ \eqref{eqn:HiggsGF}, we find the reduced action, 
\begin{align}
\mathcal{L}_{\rm red} &= \mathcal{L}_{\rm GF} +a^3 N\[ \frac{\dot\zeta^2}{2N^2}-  g^2 \zeta^2\lambda^2\frac{\phi^2}{a^2}-V(\zeta^2) \],
 \label{hybrid-Lag-red}
\end{align}
where $\mathcal{L}_{\rm GF}$ represents the terms from the first two lines of eq.\ \eqref{reduced-action}. The contributions of the Yang-Mills and $\kappa (F\tilde{F})^2$ are again given by the expressions in section \ref{sec-static}. For the Higgs-sector contribution, we can simply replace the masses $(m_1, m_2, m_3)$ with  $(0,  g\zeta,  g\zeta )$ in $\rho_m$ and $P_i^m$ from eqs.\ \eqref{rho_massive} and \eqref{p1_massive}, 
\begin{align}
\rho_{m} ~=  ~g^2 \mathbf{\rm \zeta}^2\lambda^2\frac{\phi^2}{a^2},\label{rho_massive_hy} \quad P^{m}_{1} =   - ~\mathbf{\zeta}^2\lambda^2\frac{\phi^2}{a^2}, \quad P^{m}_{2} = P^{m}_{3} = 0.
\end{align}
We also need to include the contributions to the stress tensor from the dynamical scalar
\be \label{rho_scalar}
\rho_{_{\zeta}} =   \ddfrac{1}{2} \dot\zeta^2 + V(\zeta^2), \quad 
P_{_{\zeta}} =    \ddfrac{1}{2} \dot\zeta^2 - V(\zeta^2).
\ee
The  density and pressure now read 
\bea
\rho = \rho_{_{\rm YM}} + \rho_{\kappa}+\rho_{m} + \rho_{_{\zeta}} ,\quad P_{1} = P - \frac{2}{3} \tilde{P},\quad P_{2} = P_{3} = P + \frac{1}{3} \tilde{P},
\eea
where 
\bea
P = - \rho_{\kappa} + \ddfrac{1}{3}  \rho_{_{\rm YM}}  - \ddfrac{1}{3}\rho_{m}+P_{_{\zeta}},\quad  \tilde{P} =  \tilde{P}_{_{\rm YM}}+ \tilde P_{m},\quad \rho_{m}= \tilde P_{m}   = \mathbf{\zeta}^2 g^2\lambda^2.
\eea
These quantities again lead to the three gravitational field equations, 
 \bea
&~&\dot{\alpha}^2-\dot{\sigma}^2=\frac13\( \rho_{_{\rm YM}}  + \rho_{\kappa} + g^2 \zeta^2\lambda^2 + \ddfrac{1}{2} \dot\zeta^2 + V(\zeta^2)\), \label{dot-alpha^2_hy}\\
&~&\ddot{\sigma}+3\dot{\alpha}\dot{\sigma}=\frac{P_2-P_1}{3}=\frac{\tilde P}{3}, \label{ddot-sigma_hy}\\
&~&\ddot{\alpha}+3\dot{\sigma}^2=-\frac{3\rho+P_1+2P_2}{6}=-\frac23\rho_{_{\rm YM}}-\(\ddfrac13 g^2 \zeta^2 \lambda^2 + \frac12\dot\zeta^2\)\label{ddot-alpha_hy}.
\eea

Before we analyze the anisotropic behavior, we first determine the conditions for inflation. Through (\ref{dot-alpha^2}) and (\ref{ddot-alpha}), we find the expression for $\epsilon = - \dot{H}/H^2 = - \ddot\alpha/\dot\alpha^2$, 
\be
\epsilon=\ddfrac{2\rho_{_{\rm YM}}+9\dot\sigma^2+g^2\zeta^2\lambda^2 + \frac32\dot\zeta^2}{\rho_{_{\rm YM}}+\rho_\kappa+3\dot\sigma^2 + g^2\zeta^2\lambda^2 + \ddfrac{1}{2} \dot\zeta^2 + V(\zeta^2)}\ . \label{epsilon-hy}\\
\ee
Note that, besides satisfying conditions $\rho_\kappa \gg \rho_{_YM}, \dot\sigma^2$ as in section \ref{sec-static}, we can also prolong inflation by adjusting the potential of the scalar field so that $\dot\zeta^2 \ll V(\zeta^2) \sim \rho_\kappa$. We do not consider this possibility further here, and restrict ourselves to parameters such that $V(\zeta^2) \ll  \rho_\kappa$.

\begin{figure}[t!]
\centering
 \includegraphics[width=\textwidth]{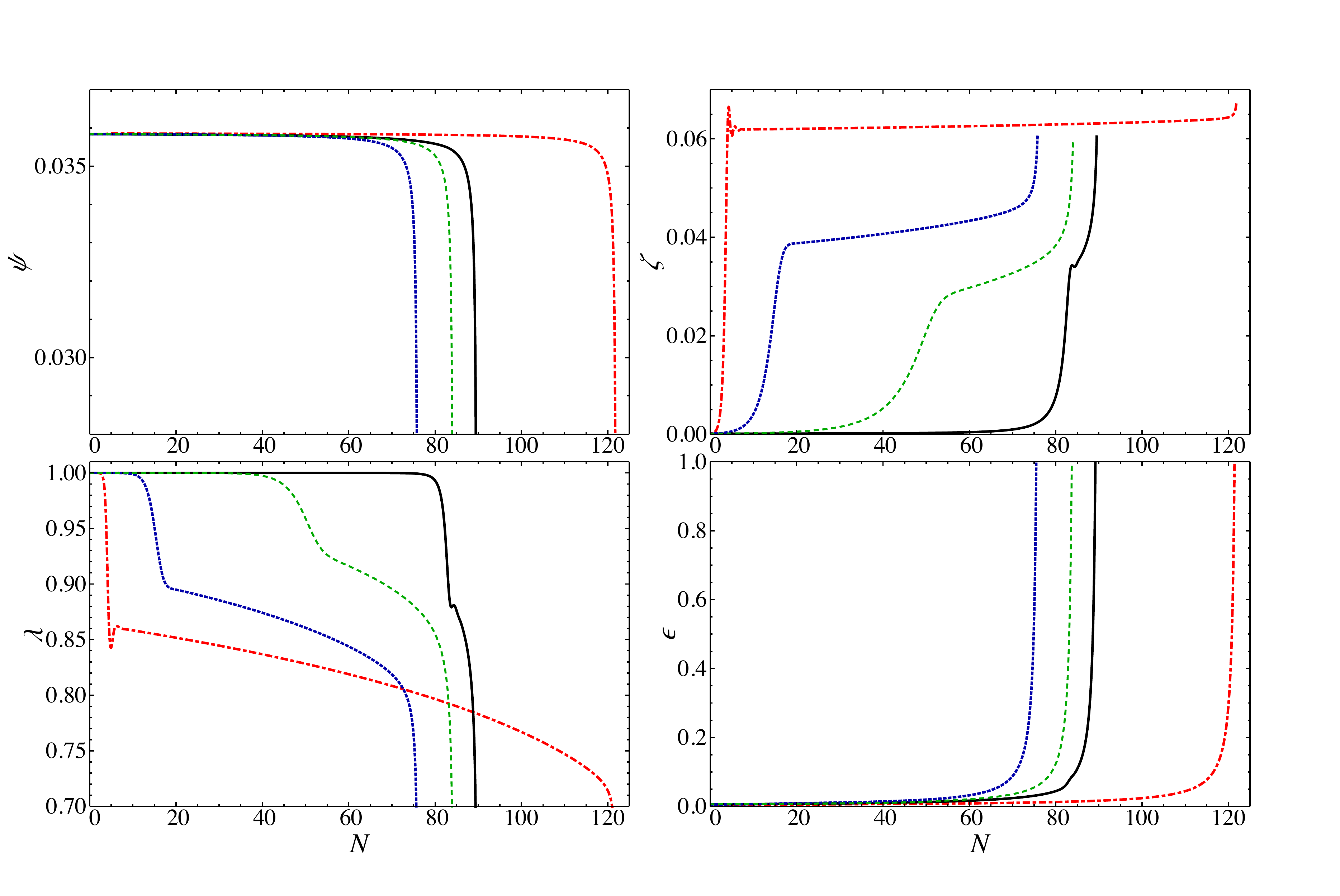}
 \caption{The full numerical evolution of dynamical symmetry breaking in massive Gauge-flation where the spacetime transition from an initially inflating quasi-de Sitter spacetime to an accelerating Bianchi spacetime. We plot $\psi(N)$ (top left), $\zeta(N)$ (top right), $\lambda(N)$ (bottom left), and $\epsilon = -\dot{H}/H^2$. To generate the curve, we choose parameters $\sqrt{\kappa} g = 30635.1$, $\psi_0 = 0.0358$, $\sqrt{\kappa} \dot\psi_0 = 10^{-10}$, $\lambda_0 = 1.0,$ $\sqrt{\kappa} \dot\lambda_0 = 1.5\times10^{-10}$, $\zeta= 10^{-4}$, $\sqrt{\kappa}\dot\zeta= 0$, $\nu= 0.097$. We vary $\mu$, and thus $\beta$ taking $\kappa \mu = 5 \times 10^8,  1.28 \times 10^8, 1.5 \times 10^8, 1.31 \times 10^8$,  corresponding to $\beta = 3.919$  (red dot-dashed curve), $\beta =  1.004$ (black solid curve) , $\beta = 1.176$ (blue dotted-curve), $\beta = 1.03$ (green dashed-curve).}\label{fig: dyn}
\end{figure}

\subsection{Evolution of the Higgs vev, and dynamical symmetry breaking during inflation}\label{eqn:dynhiggs}

The equation of motion for $\zeta$ is given by
\be
\ddot\zeta + 3\dot\alpha \dot\zeta + 2 g^2 \lambda^2 \ddfrac{\phi^2}{a^2} \zeta  +\frac{dV}{d\zeta} =0. 
\ee
For concreteness, we consider  the standard symmetry-breaking double-well shape for $\zeta$ 
\begin{align}
V = \mu\(\zeta^2- \frac{\nu^2}{2}\)^2 \, ,
\end{align}
where $(\mu,\nu)$ are parameters to be chosen. The minima of the bare potential are located at $\zeta = \pm \nu/\sqrt{2}$. However, note that the interaction with the gauge field, $2g^2\lambda^2\psi^2\zeta^2$, distorts the potential and the true minima is 
\begin{align}\label{eqn:fieldmin}
\zeta_{0} =\pm \sqrt{\frac{\nu^2}{2}- \frac{g^2\lambda^2\psi^2 }{2\mu}}.
\end{align}
Therefore, in order to break the symmetry we require  $\nu^2 > g^2 \lambda^2 \psi^2/\mu$. That is, supposing inflation starts in isotropic spacetime, i.e. $\lambda_{\rm{in}}=1$, we  require $\nu^2 > g^2 \psi_{\rm{in}}^2/\mu$. 
As in the static case, the parameter $\kappa$ can be again absorbed by rescaling the gauge coupling, $g$ and Hubble rate $H$, as before. However, the potential parameters must  also be rescaled according to $\mu \rightarrow \mu/\kappa$.  The fixed points of the motion are similar to those in the static case, with the replacement $m_1 \to 0$, $m_2 = m_3 \to g \zeta_0$, however, due to the rolling Higgs, the evolution of the system to the fixed points depends on the shape of the Higgs potential.

\begin{figure}[t!]
\begin{center}
\includegraphics[width = 0.75\textwidth]{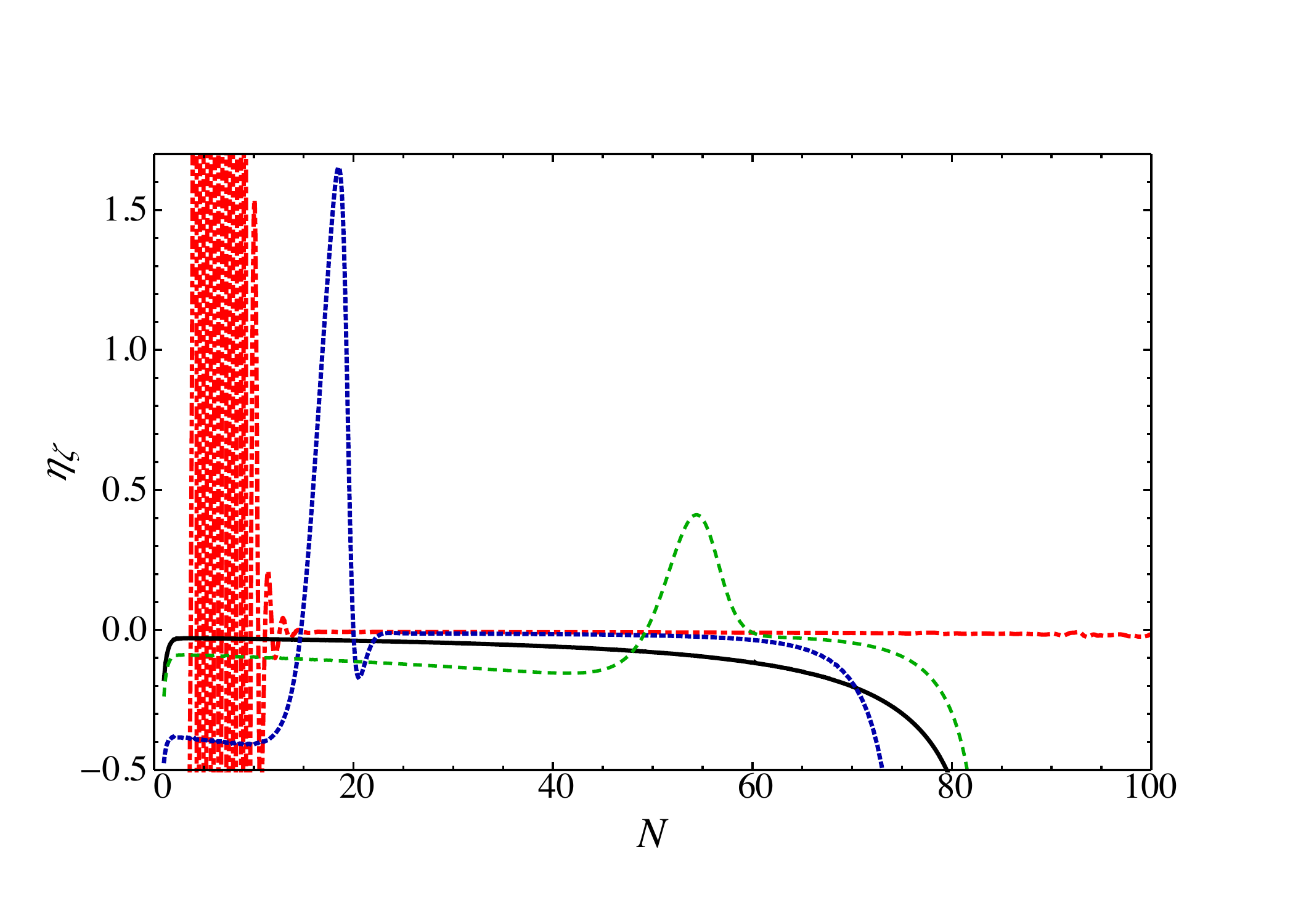}
\caption{The evolution of the slow-roll parameter $\eta_\zeta = -\ddot{\zeta}/(H\dot\zeta)$ for the symmetry breaking transitions shown in figure \ref{fig: dyn}. The curves correspond to those in figure \ref{fig: dyn}.}\label{fig:eta_zeta}
\end{center}
\end{figure}

\begin{figure}[t!]
\includegraphics[width = \textwidth]{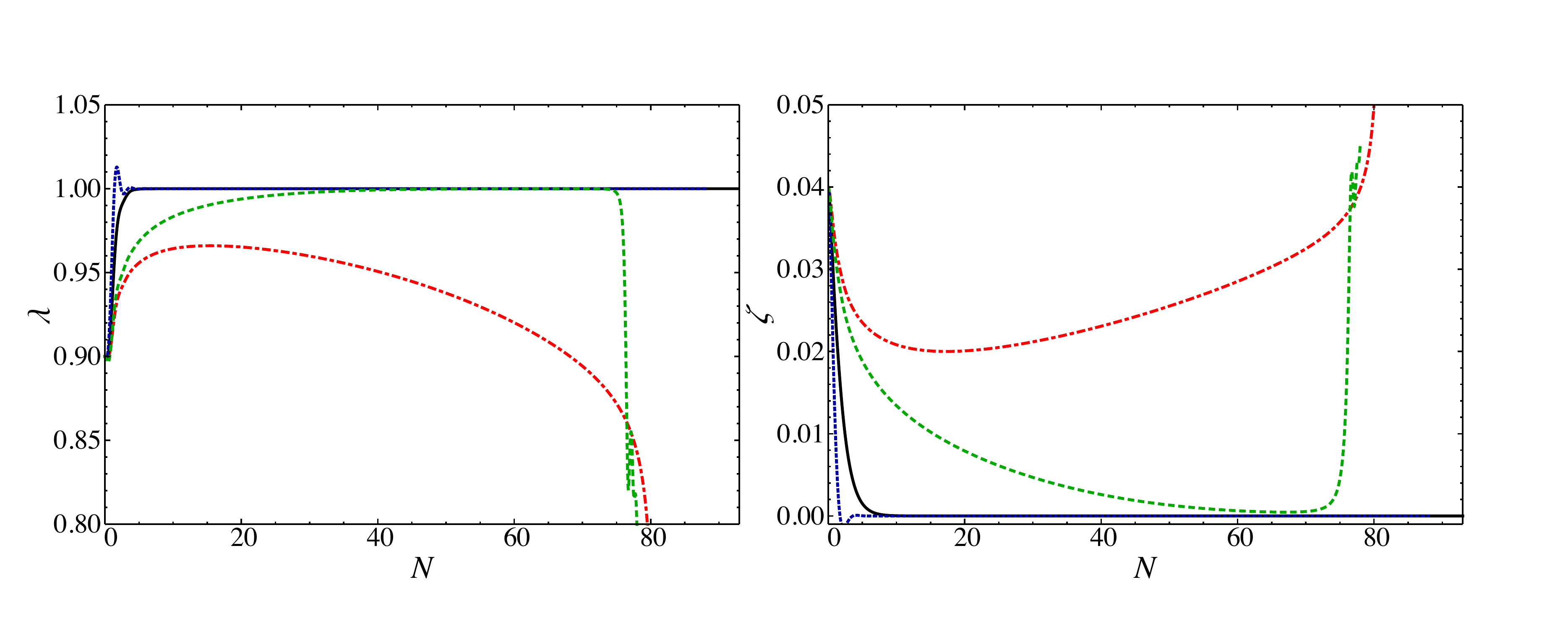}
\caption{The evolution of $\lambda$ and $\zeta$ in cases where $\beta<1$, and the symmetry is restored during inflation. In these cases, the spacetime transitions from accelerating Bianchi spacetime to quasi-de Sitter spacetime. The parameters are chosen such that  $\beta = 0.392 $ (blue dotted-curve), $\beta = 0.784 $ (black solid-curve), $\beta = 0.988$ (green dashed-curve). We also show an example with $\beta = 1.019$ (red dot-dashed curve) where initially the symmetry moves toward being restored ($\zeta = 0$, and $\lambda = 1$) before eventually breaking.} \label{fig:iso_restore}
\end{figure}

We begin by examining the conditions under which the symmetry breaking scalar $\zeta$ rolls slowly on its potential. Neglecting the acceleration term, the equation of motion for $\zeta$ becomes
\begin{align}
3H\dot{\zeta} \approx  -4\mu\(\zeta^2- \frac{\nu^2}{2}\) \zeta-2 g^2 \lambda^2 \psi^2 \zeta\, ,
\end{align}
which can be written
\begin{align}\label{eq:zeta_dot}
\frac{\dot{\zeta}}{H} \approx  -\frac{2}{3}\mu\(2\frac{\zeta^2}{\nu^2}- 1+\frac{1}{\beta}\) \frac{\nu^2}{H^2} \zeta\, ,
\end{align}
where we have defined
\be
\beta = \frac{\mu\nu^2}{(g^2 \psi_{\rm{in}}^2)} = \frac{\mu\nu^2}{\gamma H^2}.
\ee
In the limit $\zeta \ll \nu$, i.e.\ near the isotropic point,  the ratio $\beta$ controls the evolution of $\zeta$.   Note that since $\gamma \sim \mathcal{O}(1-10)$, $\beta \sim  \zeta'/\zeta$. For $\beta \gtrsim 1$, $\zeta$ slowly rolls away from the origin toward the minimum of the potential $\zeta_0$ in eq.\ \eqref{eqn:fieldmin} and the spacetime evolves from  quasi-de Sitter space to  an accelerating Bianchi spacetime (see figure \ref{fig: dyn}).  For $\beta < 1$, the symmetry will be restored as $\zeta$ evolves to the origin in field space and $\lambda$ evolves to 1 (see figure \ref{fig:iso_restore}).

Figure \ref{fig: dyn} show three possible inflation trajectories for values of $\beta >1$, where the symmetry is broken during inflation. In each example, we fix the initial values of $\{g, \psi, \dot\psi, \lambda, \dot\lambda, \zeta, \dot\zeta, \nu \}$, and vary the parameter $\mu$, and thus $\beta$, to examine different slow, or fast-roll trajectories of $\zeta$. We begin the computations with $\zeta$ near the origin in field space. Figure \ref{fig:eta_zeta} shows the slow roll parameter $\eta_\zeta = -\ddot{\zeta}/(\dot{\zeta} H)$ for each of the curves above. Note that, as expected, slow-roll ($\eta_\zeta < 1$) is not violated for the slow transition corresponding to $\beta = 1.03$, where $\zeta$ transitions to the minima over a long period.

In figure \ref{fig:iso_restore} we demonstrate that when $\beta < 1$ the symmetry is restored and an initially anisotropic spacetime transitions to quasi-de Sitter space.  For values of $\beta \lesssim1$, the isotropy restoring transition is slow, and as $\beta$ decreases toward zero, the transition occurs faster. This type of symmetry restoring transition may be interesting in the context of the large-scale anomalies in the CMB, which do not appear on smaller scales. Here, one could arrange for the transition back to isotropic expansion to happen $N\sim 50$ $e$-folds before the end of inflation.  

The curves corresponding to $\beta=1.019$ gives an instance when the restoring force combats the distorted potential. Initially, $\lambda=0.9$ gives an effective potential with a single minimum at the origin. As the $\zeta$ rolls to the origin, the spacetime evolves to become isotropic. But the increase of $\lambda$ deforms the potential and the symmetry is again broken. The universe eventually evolves towards an accelerating Bianchi spacetime. 

\subsection{Validity of the classical treatment} 

In our study of the evolution of the Higgs vev in section  \ref{eqn:dynhiggs}, we have ignored quantum fluctuations of the fields, and treated the fields as homogeneous. We now examine where this approximation is valid. 

In order to neglect the effects of quantum fluctuations, we require that the motion of the fields is dominated by their classical rolling, rather than by quantum mechanical fluctuations. We estimate the conditions under which this is true in the standard way. During one Hubble time, the Higgs vev moves a distance $\Delta\zeta \sim \dot\zeta/H$. During the same period, since it is nearly massless,  it suffers quantum fluctuations due to the near de Sitter expansion $\delta\zeta \sim H/2\pi$.\footnote{In Higgsed Gauge-flation, the fluctuations in the Higgs field mix with the fluctuations in the gauge field at linear order in perturbation theory. In general this means that approximating the field amplitude by that of a massless, free field is not a good approximation. However, it parametrically captures the amplitude of the fluctuations which is adequate for the present purposes.} For our classical treatment to be valid, we therefore require 
\be\label{eqn:cond}
\frac{\Delta\zeta}{\delta \zeta} \sim 2\pi\frac{\dot\zeta}{H^2} \gg 1.
\ee
Assuming slow-roll,  using  eq.\ \eqref{eq:zeta_dot} we can write
\begin{align}\label{eq:qm_cl_ratio}
\frac{\Delta \zeta}{\delta \zeta} \approx  -\frac{4\pi}{3}\gamma\(2\beta \frac{\zeta^2}{\nu^2}+1- \beta\)\frac{ \zeta}{H}\approx  -\frac{4\pi}{3}\gamma\(1- \beta\)\frac{ \zeta}{H}\,.
\end{align}
Recall from the discussion at the end of section \ref{sec:backgroundeqns} that the parameter $\kappa$ can be rescaled out of the dynamics.  Ultimately, $\kappa$ determines the energy scale of inflation, and thus the Hubble rate; lower values of $\kappa$ correspond to larger energy-scales and larger Hubble rates.\footnote{The value of $\kappa$ is set by matching the amplitude of the fluctuations to the observations \cite{Namba:2013kia}. This is reminiscent of the parameter $m_\phi$ in quadratic-chaotic inflation.} We are therefore always able to adjust the amplitude of the quantum fluctuations so that the condition in eq.\ \eqref{eqn:cond} is satisfied.

We can demonstrate this explicitly by rescaling eq.\ \eqref{eq:qm_cl_ratio} by $\{g, t,\mu\} \to \{g/\sqrt{\kappa}, \sqrt{\kappa}t,\mu/\kappa\}$. Since $\gamma$ and $\beta$ are invariant under this rescaling, we find
\begin{align}
\frac{\Delta \zeta}{\delta \zeta}\approx  -\frac{4\pi}{3}\gamma\(1- \beta\)\frac{ \sqrt{\kappa}\zeta}{\tilde{H}}\,,
\end{align}
where $ H = \tilde{H}/\sqrt{\kappa}$, and $\tilde{H}$ is invariant under the rescaling. Varying $\kappa$ does not affect the dynamics, and we can therefore always choose $\kappa$ such that ${\Delta \zeta}/{\delta \zeta} \gg 1$ is satisfied. In particular, when $\beta \approx 1$, i.e. when the effective potential is relatively flat, we require $\kappa$ to be larger so that the Hubble rate and the amplitude of quantum fluctuations are lower.  Note, however, that for a fixed value of $\kappa$ (which may be required by observations), the parameter $\beta$ cannot be arbitrarily close to unity.

Going beyond the classical approximation presented here and into the region where the condition in eq.\ \eqref{eqn:cond} is violated requires taking into account the quantum evolution of the Higgs field. In the case at hand, this is complicated by the fact that the Higgs interacts with the rest of the gauge field fluctuations and cannot be well-approximated as a massless free field.

An interesting possible application of the $\beta \sim 1$ limit is in the context of the curvaton scenario \cite{Lyth:2001nq,PhysRevD.67.023503, Dimopoulos:2008yv}. In the model considered here, the Higgs field could potentially behave as a curvaton field by eventually coming to dominate, or almost dominate the energy density of the Universe after inflation. During inflation, if $\beta \approx 1$, the classical Higgs will be frozen on its potential, while it will acquire fluctuations due to the usual quantum vacuum fluctuations.  As inflation ends, the gauge vev will decay, releasing the classical Higgs vev which will oscillate about the minimum of its potential. Since the gauge field redshifts like radiation after inflation, while the Higgs looks like pressureless matter, provided the Higgs is long-lived it can eventually come to dominate the energy density and act as a curvaton field.   However, because the Higgs is not a free field and the model is inherently a multifield inflation model, the details of the fluctuations require solving for all fluctuations of the gauge and Higgs field. This is a significant undertaking and is beyond the scope of this paper.

\section{Conclusions}\label{sec:conclusions}

In this work we have studied anisotropic inflationary solutions in massive Gauge-flation. The original theory of massless Gauge-flation respects the cosmic no-hair conjecture whereby initially anisotropic configurations evolve to quasi-de Sitter space within an $e$-folding. We have demonstrated that massive Gauge-flation allows for extended periods of accelerated anisotropic expansion in cases where the gauge fields acquire unequal masses. The anisotropy of the spatial slices in these cases is of the order $\Sigma \sim \mathcal{O}(\epsilon_H)$.

Working with cylindrically symmetric gauge field and spacetime configurations, we considered  dynamical symmetry breaking, as well as the case where the gauge-field masses are static, and introduced via the Stueckelberg mechanism. In the static case, by adjusting the ratio of the gauge field masses, we demonstrated that both prolate and oblate spacetime Bianchi geometries are possible. 
In the dynamical symmetry breaking case, we focused on a Higgs field that transformed under the vector representation of the gauge group. In this case, the gauge symmetry is not completely broken, and one of the gauge bosons remains massless (about the vacuum), while the other two acquire equal masses.  We demonstrated that symmetry breaking can occur quickly, with only a small (large) amount of isotropic expansion followed by an extended (short) anisotropic phase, or more slowly where the Higgs slowly rolls over many $e$-foldings to its effective minima. Our study was restricted to the vector representation for the Higgs field; it would be interesting to explore the evolution in different representations.

In the scenario where the Higgs dynamically evolves on its potential, we have restricted ourselves to the region of parameter space where the evolution of the Higgs is dominated by its classical roll. Since the overall energy scale of inflation in this scenario is a free parameter, the amplitude of quantum fluctuations can be adjusted independently of the dynamics. This freedom guarantees that there is always a region of parameter space where our analysis is valid. It would be interesting to study the opposite limit in which the Higgs evolves stochastically on its potential, we leave this to future work.

Finally, it would be interesting to study the spectra of scalar curvature, and gravitational wave fluctuations about the anisotropic attractor solution, as well as during the dynamical evolution from the accelerating Bianchi spacetime back to the isotropic quasi-de Sitter phase. We anticipate that this will be a rather complicated undertaking, and leave it to future work.

\acknowledgments This work was supported in part by the US Department of Energy through grant DE-SC0015655.

\bibliographystyle{JHEP}
\bibliography{Ani_MGF}

\end{document}